# A ridesharing simulation model that considers dynamic supply-demand interactions

**Rui Yao, Shlomo Bekhor***

Department of Civil and Environmental Engineering,

Technion – Israel Institute of Technology, Haifa 32000, Israel

Email:  andyyao@campus.technion.ac.il,

　　　　sbekhor@technion.ac.il

* Corresponding author

**Abstract**

This paper presents a new ridesharing simulation model that accounts for dynamic driver supply and passenger demand, and complex interactions between drivers and passengers. The proposed simulation model explicitly considers driver and passenger acceptance/rejection on the matching options, and cancellation before/after being matched. New simulation events, procedures and modules have been developed to handle these realistic interactions. Ridesharing pricing bounds that result in high matching option accept rate are derived. The capabilities of the simulation model are illustrated using numerical experiments. The experiments confirm the importance of considering supply and demand interactions and provide new insights to ridesharing operations. Results show that higher prices are needed to attract drivers with short trip durations to participate in ridesharing, and larger matching window could have negative impacts on overall ridesharing success rate. Comparison results further illustrate that the proposed simulation model is able to replicate the predefined "true" success rate, in the cases that driver and passenger interactions occur.

# 1. Introduction

Innovative shared mobility, namely ride-sourcing services, have reshaped our urban transportation. Ride-sourcing companies, such as Uber, Lyft, and Didi, provide convenient mobility services with lower fares, by utilizing drivers' own vehicles instead of company fleets to provide services (Wang and Yang, 2019). One type of these services is ridesharing, in which peer drivers serve more than one passenger in each ride, and typical examples of ridesharing include Grab Hitch and Didi Hitch. Ridesharing services potentially can reduce vehicles kilometer traveled (VKT), compared to other service types. For example, ride-hailing services have been criticized for intensifying the urban traffic congestions (Clewlow and Mishra, 2017; Erhardt et al., 2019; Diao et al., 2021). This is because peer drivers in ridesharing, who are assumed to perform activities other than only pick up and drop off passengers (as is the case of dedicated drivers), have their designated destinations and do not cruise in the network (Wang and Yang, 2019).

In general, a successful design of ridesharing systems is expected to alleviate traffic congestions and reduce air pollutions. However, from planning perspective, designing and evaluating the effectiveness (expected "real-world" performance) of ridesharing systems is challenging in multiple aspects, as outlined in the following paragraphs.

In a dynamic ridesharing system, on-demand driver and passenger requests for ridesharing may not be known in advance (compared to traditional carpooling). Service providers should be able to dynamically match together potential ridesharing drivers and passengers, who may enter the ridesharing system at different times. Furthermore, a driver with matched/on-board passengers (from previous matching rounds) may be matched again with new passengers (Agatz et al., 2012). Alonso-Mora et al. (2017) propose a general many-to-many dynamic multi-passenger vehicle assignment framework based on the idea of shareability networks (Santi et al., 2014), and show their algorithm's ability and efficiency of handling large demands. Simonetto et al. (2019) simplify the ridesharing matching problem and solve it efficiently using linear programming. The dynamic features of a ridesharing system could improve the system efficiency but bring additional mathematical complexity for finding matches between ridesharing drivers and passengers.

Moreover, in a realistic ridesharing setting, peer drivers and passengers may make various decisions and actions with respect to the dynamic ridesharing services. For example, as new passenger requests continue to emerge during ridesharing, existing schedules might be modified to serve these new passengers, which means the actual realization of passenger trips are related to other matched passengers (Fielbaum and Alonso-Mora, 2020). Consequently, for passengers already waiting for pickups, they may cancel their trips due to delayed pick-up after being assigned a driver (He et al., 2018; Wang et al., 2020). Similarly, for drivers and on-board passengers, they may reject the assigned matchings due to extra detours (Chu et al., 2018; Rosenblat et al., 2017). Other decisions could be passengers cancelling their trips because of a long wait before being matched with a driver (Wei et al., 2020), and passenger reordering after cancellation. In terms of evaluating a ridesharing system, if these possible driver and passenger actions were not considered in the ridesharing models, the system performance could not be properly estimated.

In addition, pricing scheme, in which ridesharing drivers are compensated by the passengers for the shared rides, could also affect matching acceptance/rejection decisions. Several pricing schemes exist in the literatures. For example, Liu and Li (2017) proposed a pricing scheme with the aim to reduce traffic congestions. A reward scheme integrated with surge pricing was proposed by Yang et al. (2020) to balance ridesharing supply and demand. Li et al. (2020) derived ridesharing price intervals such that there exists equilibrium with a positive ridesharing ridership. However, ridesharing pricing bounds have not been derived with explicit considerations of matching acceptance/rejection, and with the goal to have more drivers and passengers accepting the matchings.

The studies above indicate that the design and evaluation of a ridesharing system should include these dynamic supply-demand interactions. A considerable amount of research effort has been placed to model ridesharing systems, which could be broadly divided into analytical and simulation models. The ridesharing system is modeled analytically as a network assignment problem, or as a market equilibrium problem. Analytical ridesharing traffic assignment models are typically modeled on a supernetwork with duplicated links representing ridesharing services (Xu et al., 2015; Di and Ban, 2019; Ma et al., 2020), with matching accept/reject modeled as a spatial equilibrium problem (Li et al., 2020), and extended to account for cancellation before being matched in a time-dependent network (Wei et al., 2020). For the market equilibrium modeling approach, ridesharing systems are modeled in an aggregated context without considering network structures (He et al., 2018; Wang et al., 2020). However, most of these analytical models emphasized the (long-term) equilibria instead of time-varying operations of ridesharing systems.

To cope with the time-dependent dynamics of ridesharing and consider the complex interactions between drivers and passengers, simulation models were developed. For example, Djavadian and Chow (2017a, b) modeled the ridesharing system as an agent-based equilibrium problem with a day-to-day learning process. Wang et al. (2017) implemented an agent-based dynamic ridesharing model in the MATSim platform (Horni et al., 2016), with explicit consideration that drivers can switch their roles as drive alone if they would be late for their next activities. Beojone and Geroliminis (2021) proposed a simulation framework that accounts for passenger cancellation before assigning a driver. Nahmias-Biran et al. (2019) considered ride-hailing services in SimMobility simulation framework (Adnan et al., 2016; Oke et al., 2020), in which dedicated driver agents consider stopping service and leave the platform. In Shen et al. (2018), passengers may consider leaving shared mobility system if waiting too long for the service. Thaithatkul et al. (2019) developed an agent-based ridesharing simulation which considers passenger acceptance/rejection on peer passengers. Conversely, Martinez et al. (2015) and Linares et al. (2016) considered passenger acceptance/rejection on drivers. The driver and passenger acceptance/rejection behavior were also captured by an auction matching process in Nourinejad and Roorda (2016). A summary of existing ridesharing models is given in Table 1.

Table 1 A comparison between the proposed and existing ridesharing models.

| Literature | Dynamic supply demand | Multiple passengers | Cancellation before being matched | Matching option acceptance/rejection | Cancellation after being matched |
|---|---|---|---|---|---|
| **Network assignment models** | | | | | |
| Xu et al. (2015) | | √ | | | |
| Di and Ban (2019) | | √ | | | |
| Ma et al. (2020) | | √ | | | |
| Li et al. (2020) | | | | √ | |
| Wei et al. (2020) | √ | | √ | | |
| **Market equilibrium models** | | | | | |
| He et al. (2018) | | | | | √ |
| Wang et al. (2020) | | | | | √ |
| Ke et al. (2020) | | √ | | | |
| **Simulation models** | | | | | |
| Djavadian and Chow (2017a, b) | √ | √ | | | |
| Wang et al. (2017) | √ | √ | √ | | |

| Literature | Dynamic supply demand | Multiple passengers | Cancellation before being matched | Matching option acceptance/rejection | Cancellation after being matched |
|---|---|---|---|---|---|
| Beojone and Geroliminis (2021) | √ | √ | √ | | |
| Shen et al. (2018) | √ | | √ | | |
| Nahmias-Biran et al. (2019) | √ | | | | |
| Thaithatkul et al. (2019) | √ | | | √ | |
| Linares et al. (2016) | √ | √ | | √ | |
| Nourinejad and Roorda (2016) | √ | √ | | √ | |
| **This paper** | √ | √ | √ | √ | √ |

As shown in the table, there are still gaps in the literature for developing a comprehensive ridesharing simulation model, with explicit considerations of all the possible actions drivers and passengers may take during ridesharing services.

*1.1. Objectives and Contributions*

We propose to develop a ridesharing simulation model that considers the complex dynamic driver supply and passenger demand interactions. More specific, our ridesharing simulation model can handle drivers' and passengers' arrivals at different times in the system, their acceptances/rejections on different matching options, and both driver and passenger order cancellation and reordering at different matching stages, within one unified framework.

As discussed before, these realistic actions are important for understanding the effect of services qualities on driver supplies and passenger demands. Our simulation model provides not only aggregated metrics of a ridesharing system, but also realistic agent behaviors. This is important for developing and evaluating more refined operational strategies, for example, personalized ridesharing trip planning, or dynamic pricing, which are expected to improve the system efficiency. Furthermore, our simulation model provides additional modeling choices regarding how acceptance/rejection, cancellation, and alternative mode decisions are simulated, these models capture the behavioral aspects of drivers and passengers in ridesharing and are expected to provide additional insights for developing more effective policies.

In the context of simulation model, new simulation events, procedures, and modules are developed to capture the complex dynamic interactions, to provide new insights, and to handle multiple realistic scenarios. The contributions of this paper are then summarized as follows:

- It proposes a new simulation model with abilities to handle dynamic driver supply and passenger demands, driver and passenger acceptance/rejection on matching options, and order cancellation before/after being matched, altogether.
- It provides insights on the detailed time-varying dynamic process of a ridesharing system under more realistic settings, and the overall system performance at the network level.
- It offers a comprehensive and flexible testbed for designing and evaluating more refined operational strategies and policies with the new abilities to capture matching option acceptance/rejection, and order cancellation behaviors. Specifically, bounds on ridesharing pricing scheme, which could result in high matching option accept rate, are derived and evaluated using the proposed simulation model.
- It brings new choice modeling options regarding driver and passenger decisions on matching option acceptance/rejection, order cancellation, and alternative modes, and these models are implicitly integrated in our simulation model.

## 2. Methodology

The proposed ridesharing simulation model is consisted of 4 main components: 1) the simulation core; 2) the service provider; 3) the ridesharing passenger agent; and 4) the ridesharing driver agent (Figure 1). As indicated in the previous section, the connections between these components are explicitly designed to handle a) Dynamic passenger demand and driver supply; b) Acceptance/rejection on matching options; c) Order cancellation before/after being matched.

The proposed ridesharing simulation architecture resembles the general two-sided structure of a ridesharing system, in which the service provider enables interactions between ridesharing drivers and passengers (as indicated in the bidirectional connections in Figure 1).

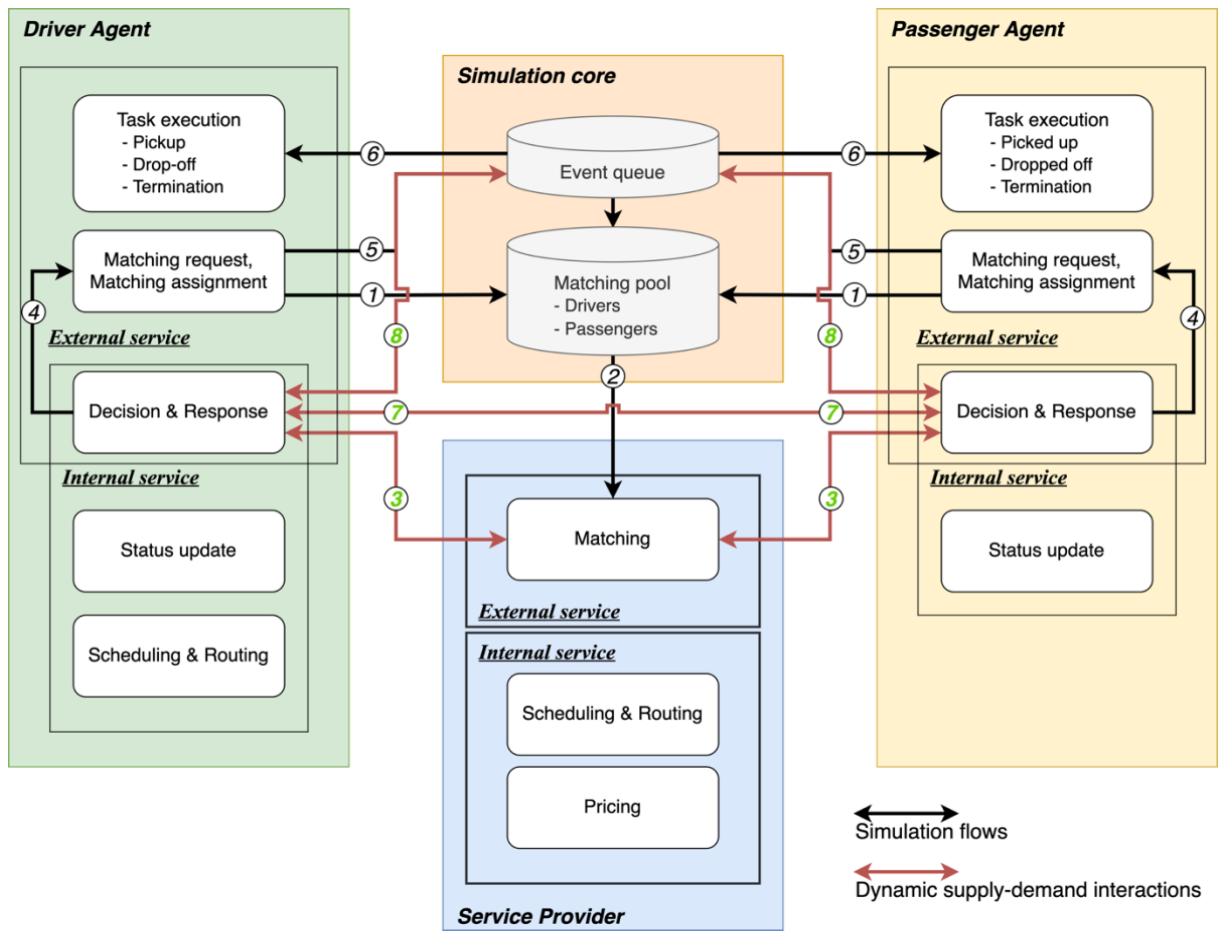

Figure 1 Ridesharing simulation architecture

The simulation model attempts to mimic realistic ridesharing scenarios, in which drivers and passengers specify their origins and destinations with their desired departure times, time constraints (e.g., maximum driving/riding time, latest arrival time), the number of seats needed (for passengers), the vehicle capacities (for drivers), and the associated service providers. These ridesharing drivers and passengers send their requests and wait for the matchings (*Connection 1*).

The simulation model handles the batched ridesharing requests to one or more service providers. These providers then match drivers and passengers according to their requested information (*Connection 2*). The matching options, which includes matching information (e.g., estimated pickup times, riding times, number of peer passengers and earnings/costs), are continuously updated to both drivers and passengers. They in turn can decide whether to accept the matching options (*Connection 3*). The ridesharing service providers then process these acceptance/rejection responses and finalize the matchings according to their operational strategies. These finalized matchings are assigned to drivers and passengers (*Connection 4*), and the drivers schedule to serve the matched passengers (*Connection 5*) and perform pickups and drop-offs (*Connection 6*).

In a realistic ridesharing scenario, drivers and passengers may cancel their orders if they think the ridesharing service does not satisfy their travel needs anymore. For example, the pickup times could increase due to traffic congestion or speed overestimates. Consequently, it may be too late for passengers (or drivers) to reach their destinations, and they might need to cancel their orders (*Connection 7*). Moreover, if the matching window is too long, drivers and passengers may cancel their requests before being matched (*Connection 8*). The following subsections further explain the four main boxes in Figure 1: 1) simulation core; 2) service provider; 3) passenger agent; and 4) driver agent. For interested readers on additional implementation details of the simulation model, we refer to the supplementary materials.

*2.1. Simulation core*

Our proposed ridesharing simulation is an event-based time-dependent model, in which an event queue is designed to handle events in a chronological order. The events can be grouped into 3 main categories as follows:

1) Ridesharing events: agent pickups, drop-offs and reaching destinations (termination), which send ridesharing drivers to serve the matched passengers, and remove the driver and passenger agents from the simulation once they reach their destinations.

2) Agent decision-making events, which correspond to the order cancellation behaviors before/after being matched.

3) Service provider matching events, which match a batch of drivers and passengers, and handle the matching option acceptance/rejection.

We assume a ridesharing service that batches new driver and passenger requests for every specific time interval, and the service providers perform ridesharing matching at the end of the time intervals. These new ridesharing requests are recorded in the matching pool and sent to service providers for matchings. We further assume that passenger agents, who either are matched or decide to quit ridesharing (e.g., due to long waiting times), will not participate in future matchings (i.e., they will be removed from the matching pool). Conversely, we allow drivers, who have been matched with or without onboard passengers, to be matched again with new passengers, unless they decide to quit ridesharing.

## 2.2. Service provider

The main functionalities of the service provider are to, 1) match ridesharing drivers and passengers in the matching pool; and 2) set prices for the matching options.

Ridesharing matching involves solving a lower-level vehicle routing problem (VRP), and an upper-level matching problem. The lower-level VRP is solved for each ridesharing driver $v_i$ and a given set passengers $\{r_j, \dots, r_k\}$ (both candidate and already matched passengers, if any). Given the spatiotemporal constraints of the ridesharing requests (e.g., maximum driving/riding time, latest arrival time), capacity constraints, and conservation constraints (e.g., drop off after pickup), the VRP is solved for the optimal pickup and drop-off sequence for the given set of passengers, while satisfying the constraints. An efficient dynamic tree algorithm (Yao and Bekhor, 2021) is adapted in this paper for solving the VRP. The fundamental idea of this algorithm is to keep track of the feasible VRP solutions for a given set of passengers using a dynamic tree structure, in which new feasible solutions can be obtained by inserting new passenger requests into the existing tree. For example, given the exiting dynamic tree of driver $v_1$ with previously matched passengers $r_1$ (Figure 2a), the new passenger request $r_2$ is inserted by traversing the tree (Figure 2b), and the optimal sequence can be obtained. For further details of the algorithm, we refer to Yao and Bekhor (2021).

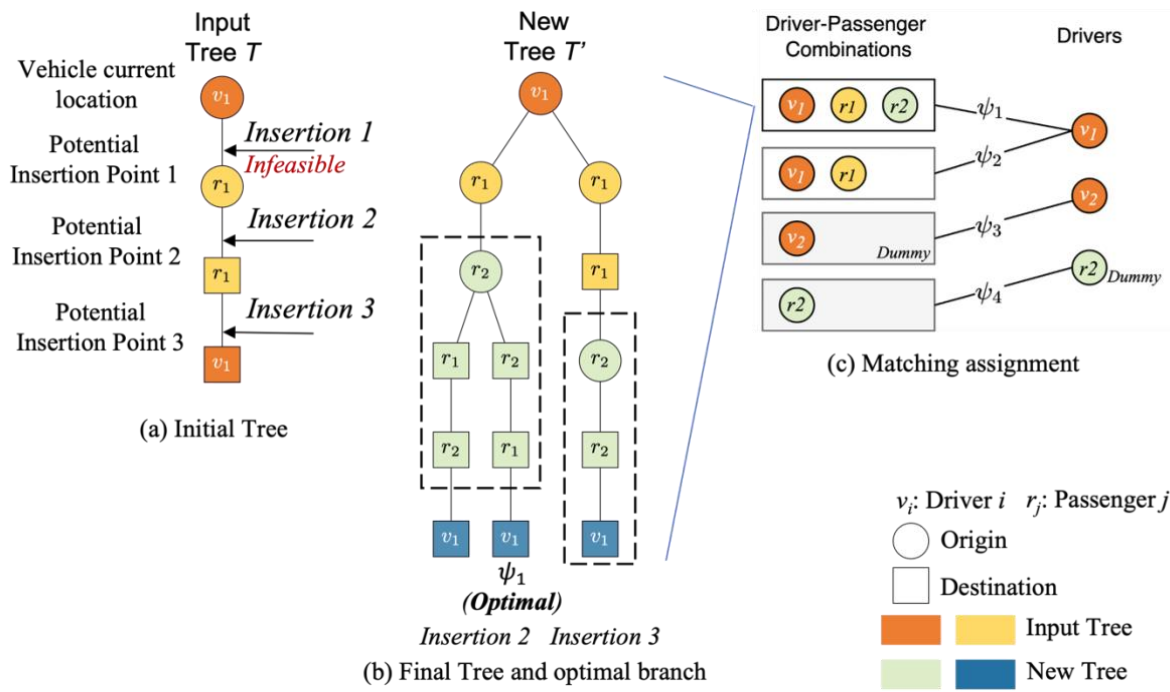

Figure 2 Ridesharing matching (adapted from Yao and Bekhor, 2021)

The upper-level matching problem is formulated as a bipartite matching problem, where one side of the graph is the drivers, and the other side is the feasible driver-passenger combinations (Figure 2c). These combinations are generated by inserting new passengers into feasible driver-passenger combinations (incrementally in size) (Alonso-Mora et al., 2017), and their feasibilities are verified by the lower-level VRP. For every feasible driver-passenger combination $m$, there will be a link with weight $\psi_m \in \Psi$, connecting the corresponding driver. In addition, a dummy driver (or dummy combination) is created for each passenger (or unmatched driver), which represents the case that passenger (e.g. $r_2$) (or driver $v_2$) is not matched in this matching window and associated with a link weight equal to the traveling alone.

We assume a public multi-passenger ridesharing system, which has a societal objective to provide a convenient ridesharing service and reduce the total vehicle kilometer traveled (VKT). In this case, link weights $\psi_m$ represent the VKT of matching $m$. The ridesharing matching problem assigns a set of drivers $V$ and a set of passengers $R$ to the driver-passenger combinations $M$.

The mathematical formulation is defined as follows:

$$\min z = \Psi \cdot \Delta \tag{1}$$

subject to:

$$\sum_{m=1}^{|M|} \omega(j,m) \cdot \delta(m) \leq 1, \forall j \in [1, |R|] \tag{2}$$

$$\sum_{m=1}^{|M|} \tau(i,m) \cdot \delta(m) \leq 1, \forall i \in [1, |V|] \tag{3}$$

$$\delta(m) \in \{0,1\}, \forall m \in [1, |M|] \tag{4}$$

Where, $\Psi$ is a vector of VKTs of $|M|$ feasible driver-passenger combinations, and vector $\Delta$ is the decision variable, in which each element $\delta(m) \in \{0,1\}$. If a driver-passenger combination is matched then $\delta(m) = 1$, otherwise $\delta(m) = 0$. The passenger-combination incident matrix is defined as $\Omega$ of size $[|R|, |M|]$, in which element $\omega(j,m)$ is equal to 1 if the passenger $j$ is included in driver-passenger combination $m$. The driver-combination incident matrix is defined as T of size $[|V|, |M|]$. If driver $i$ is associated with driver-passenger combination $m$, then $\tau(i,m) = 1$, otherwise $\tau(i,m) = 0$.

After solving the matching problem, the cost-earnings are calculated. The pricing scheme is adapted from Agatz et al. (2011), in which the variable costs of the joint trip are proportional allocated based on the distances of the separate trips, and the driver's earning is equal to the total costs (i.e. assume no commission fees in a public ride-sharing setting). We illustrate the cost allocation of matching option $(v_1, r_1, r_2)$ in Figure 3, and an example of the matching options is provided in the supplementary materials (SI Table 3, and 4).

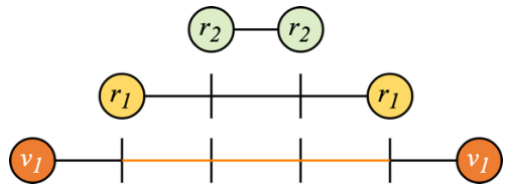

Figure 3 Example of cost allocation

Furthermore, we assume drivers and passengers only decide on accept/reject new passengers, and do not change their previously agreed matching at this stage (the changes in previously agreed matching will be explicitly handled as order cancellations). After receiving the responses from the drivers and passengers, the service provider finalizes these matching options (e.g., by majority voting, see supplementary material), and assign matchings to the corresponding drivers and passengers.

*2.3. Passenger agent*

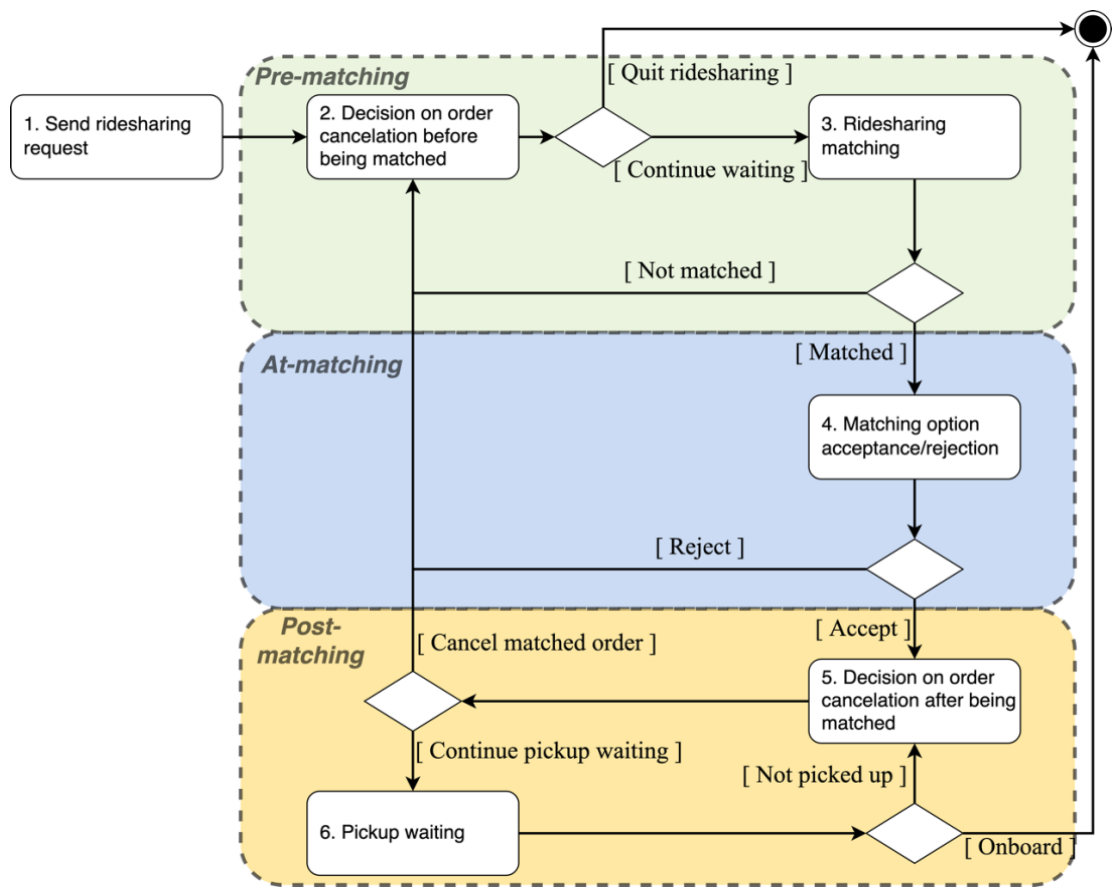

Figure 4 Passenger agent main activities

The passenger agent main activities in ridesharing can be defined by three stages: pre-matching; at-matching and post-matching (Figure 4), in which ridesharing decisions are continuously made by passenger agents regarding order cancellations and matching option acceptance/rejection.

A passenger agent sends his/her requests for a shared ride and starts waiting for a matching. In a dynamic ridesharing that accounts for agent order cancellations and re-ordering, passengers may request for matchings several times. As illustrated in Figure 5, we define the waiting time $wt(r_j)$ from the moment a passenger sends his/her 1st request to account for re-ordering. We further assume that passengers evaluate their utilities between traveling alone now (i.e., quit ridesharing), and continuing with ridesharing. Note that the proposed simulation model is general, in the sense that it can accommodate different decision models (i.e., utility functions, and choice model structures). For illustration purposes, we define in the following subsections simple utility functions with two key components: (waiting and travel) times and costs, for passenger decision makings at each stage. Alternatively, more sophisticated utility functions can be used to capture additional trade-offs (e.g., time-reliability-cost trade-offs as in Alonso-Gonzalez et al., 2020).

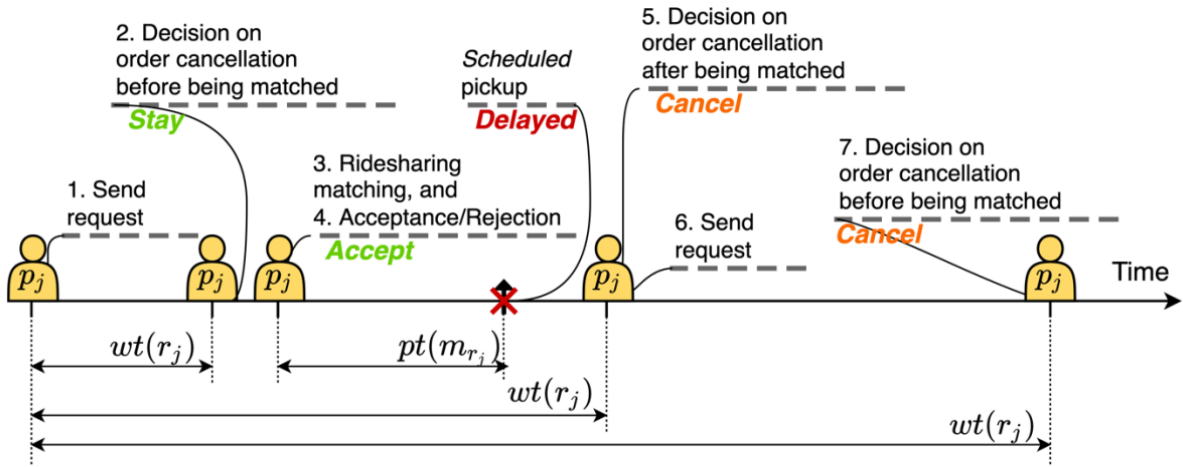

Figure 5 Example of passenger events and waiting times

*2.3.1. Pre-matching: Order cancellation before being matched*

As illustrated in Figure 5, passenger agents may decide several times on whether continue waiting for matching (i.e., stay in ridesharing), or cancel their orders. This can happen before the first matching (e.g., event 2 in Figure 5), or after re-ordering (e.g., event 7) due to cancellation of an already matched order. We also assume that passengers quit ridesharing and travel alone, if they cancel orders before being matched, and they cancel orders if they have higher utilities.

The utility of cancelling orders before being matched for passenger $r_j$, $U_{r_j,0}$, is defined equivalently as the utility of traveling alone:

$$U_{r_j,0} = \beta_{time,r_j} \cdot tt_r(r_j) + \beta_{cost,r_j} \cdot cost(r_j) \tag{5}$$

where, $\beta_{time,r_j}$ is the coefficient for travel time and $\beta_{cost,r_j}$ is coefficient for costs. And $tt_r(r_j)$ and $cost(r_j)$ represent the travel time and cost if passenger travel alone on the shortest path from his/her origin to the destination.

The utility of passenger $r_j$ staying in ridesharing system is specified by equation (6):

$$U_{r_j,1} = \beta_{time,r_j} \cdot [tt_r(r_j) + wt(r_j)] + \beta_{cost,r_j} \cdot [\beta_{\text{exp pay},r_j} \cdot cost(r_j)] \tag{6}$$

where, $wt(r_j)$ is the waiting time as discussed above, and $\beta_{\text{exp pay},r_j}$ represents passenger's expectation on the trip cost. The first part of the utility represents the total trip time if the passenger could be matched now with a direct ride. The second part represents the agent's experiences on the possible payment. The need of payment perception is due to the cost allocation scheme (Figure 3), in which the cost of a shared segment is split between passengers and known only after matching.

*2.3.2. At-matching: Matching option acceptance/rejection*

Once service providers complete the matching, the matching options are provided to the corresponding passenger agents. We assume each agent chooses one of these matching options, which also include the existing matching option (if already matched) or staying alone option (if not matched yet). Furthermore, acceptance/rejection decisions are probabilistically simulated, rather than deterministic.

The utility of matching option $m$ for passenger $r_j$, $U_{r_j,m}$, is defined as follows:

$$U_{r_j,m} = \beta_{time,r_j} \cdot [tt_r(m_{r_j}) + pt(m_{r_j})] + \beta_{cost,r_j} \cdot cost(m_{r_j}) \tag{7}$$

where, $\beta_{time,r_j}$ is the coefficient for time. And $tt_r(m_{r_j})$ and $pt(m_{r_j})$ represent the travel time and scheduled pickup waiting time of matching option $m$ for passenger $r_j$, provided by the service provider. Note that, as illustrated in Figure 5, after being matched, $pt(m_{r_j})$ is only one portion of the total waiting time $wt(r_j)$. We assume that the decisions on matching option acceptance/rejection are referred to the moment received matching, whereas cancellation decisions due to long matching waiting times or delayed pickups are handled at the pre/post-matching stages, and the decisions can be realized from various choice models, for example, multinomial logit (MNL):

$$P_{r_j}(m) = \frac{e^{U_{r_j,m}}}{\sum_n e^{U_{r_j,n}}} \tag{8}$$

*2.3.3. Post-matching: Order cancellation after being matched*

After matching, passengers, who accepted the matching options, wait for pickups. We assume that passengers stay with their matchings if there is no change in 1) scheduled pickup waiting time, 2) scheduled trip time, and 3) estimated cost of the matching options. If either of these three components changes, passengers can decide whether to cancel the already matched order (if not yet picked up). Note that, the changes in these components could be caused by delays (e.g., speed overestimates by the service provider), or cancellation of other agents (by either driver or passenger agents), which triggers a recursive decision making. We refer interested readers for detailed implementations of this recursive decision making to the supplementary materials.

We define the utility of cancelling orders after being matched for passenger $r_j$, $U_{r_j,2}$, as follows:

$$U_{r_j,2} = \beta_{time,r_j} \cdot tt_r(r_j) + \beta_{cost,r_j} \cdot [cost(r_j) + CancellationFee] \tag{9}$$

Note that, $U_{r_j,2}$ is similar to $U_{r_j,0}$ (cancellation before being matched), except for the additional $CancellationFee$, which discourages passengers to cancel already matched orders.

The utility of passenger waiting for the scheduled pickup of matching option $m$ is specified as $U_{r_j,3}$, by equation (10):

$$U_{r_j,3} = \beta_{time,r_j} \cdot [tt_r(m_{r_j}) + wt(r_j)] + \beta_{cost,r_j} \cdot cost(m_{r_j}) \tag{10}$$

Utility function (10) is very similar to $U_{r_j,1}$ (staying in ridesharing before being matched), with only the travel time and cost to be the scheduled value instead of passenger's assumptions. The $wt(r_j)$ term is added to capture the behavior of order cancellation due to long pickup waiting times.

## *2.4. Driver agent*

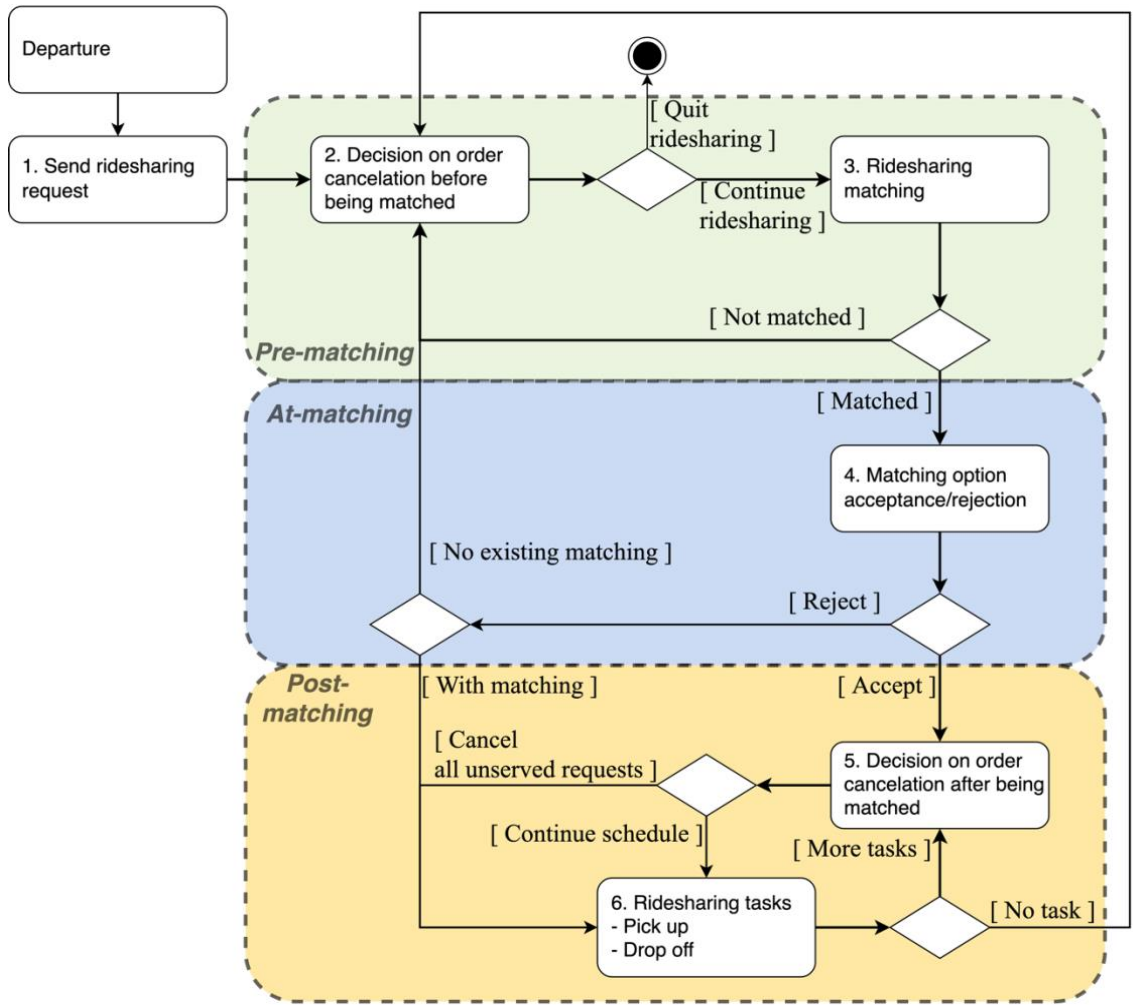

Figure 6 Driver agent main activities

Similar to passenger agents, the main activities of driver agents can also be defined by pre-matching, at-matching and post-matching stages (Figure 6). One major difference between them is that driver agents are traveling continuously to either pick up, drop off, or reach their destinations (termination), while passenger agents wait at their origins for pickups.

A driver agent departs from his/her origin (to his/her destination), while ridesharing requests may occur. We assume an on-demand ridesharing setting, in which drivers continuously receive (new) matchings (events 3, 5 in Figure 7) and could deviate from their current route for picking up new passengers. The simulation model keeps track of the location and remaining seats of ridesharing drivers, from the moment they depart.

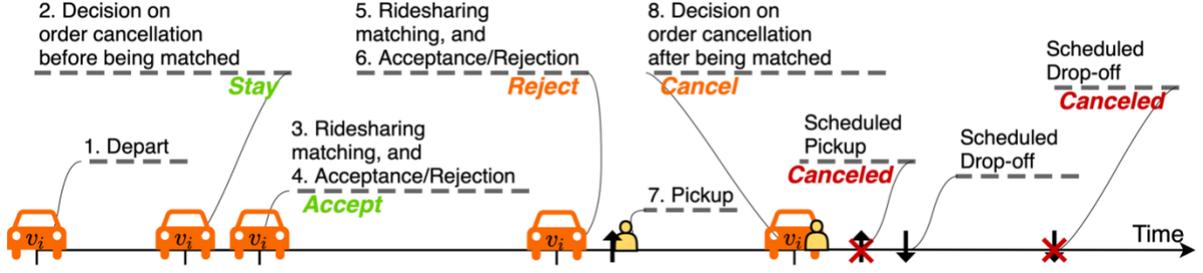

Figure 7 Example of driver events

Driver agents then make ridesharing decisions regarding order cancellations and matching option acceptance/rejection while traveling. As illustrated in Figure 7, we assume that if driver agents reject new matchings (event 6), they continue their existing schedules (e.g., event 7, to pick up previously matched passenger). Furthermore, if there is any onboard passenger, the driver should deliver these passengers. This means if drivers choose to cancel the matched orders, he/she can only cancel unserved passenger requests (i.e., matched but not yet picked up). If drivers continue their schedules, he/she could decide again (later) whether to cancel the matched orders. In the following subsections, we specify the utility functions for these driver ridesharing decisions in the three stages, respectively.

*2.4.1. Pre-matching: Order cancellation before being matched*

As shown in Figure 6, driver agents can decide several times on whether staying in ridesharing services or quitting the service. This decision making can happen even before the driver has been matched (e.g., event 2 in Figure 7), or after delivering all previously matched passengers. We assume that, if driver agents quit ridesharing, they will travel directly to their destinations, and consequently will be removed from the simulation. Furthermore, it is assumed that drivers choose the option (staying vs. quitting) with the highest utility.

The utility of cancelling orders before being matched (or when no more passengers to be delivered) for driver $v_i$, $U_{v_i,0}$, is defined equivalently as the utility of traveling alone:

$$U_{v_i,0} = \beta_{time,v_i} \cdot tt_v(v_i) + \beta_{cost,v_i} \cdot cost(v_i) \quad (11)$$

where, $\beta_{time,v_i}$ and $\beta_{cost,v_i}$ are the coefficients for travel time and cost, respectively. And $tt_v(v_i)$ and $cost(v_i)$ represent the travel time and cost if driver travel alone on the shortest path from his/her current location to the destination.

And the utility of driver $v_i$ staying in ridesharing system is defined as:

$$U_{v_i,1} = \beta_{time,v_i} \cdot [tt_v(v_i) + \Delta t(v_i)] + \beta_{cost,v_i} \cdot [cost(v_i) - \beta_{\exp\ earn,v_i} \cdot earning(v_i)] \quad (12)$$

where, $\Delta t(v_i)$ is the remaining excess travel time (such that the driver will not be late, and $\Delta t(v_i)$ = max excess travel time the moment he/she departs), $earning(v_i)$ represents the payment received from the passengers if they travel with the driver the whole trip, and $\beta_{\text{exp earn},v_i}$ represents driver's expectation on the earning (i.e., percentage of billable distances). We further assume that, if $\Delta t(v_i) = 0$, drivers quit ridesharing and travel directly to their destinations, since it could be burdensome to pick up additional passengers even without the need to detour.

*2.4.2. At-matching: Matching option acceptance/rejection*

Similar to the passenger agents, driver agents also choose between new matching options, and their existing matching option (if already matched) or staying alone option (if not matched yet). Their decisions on the matching options are also probabilistically simulated.

The utility of matching option $m$ for driver $v_i$, $U_{v_i,m}$, is defined as follows:

$$U_{v_i,m} = \beta_{time,v_i} \cdot tt_v(m) + \beta_{cost,v_i}[cost(m) - earning(m)] \qquad (13)$$

where, $tt_v(m)$ is the scheduled travel time from the current location until the driver's destination. $cost(m)$ is the estimated operational cost, and $earning(m)$ is the estimated total payment received from the passengers. Note that, utility function (13) is similar to $U_{v_i,1}$ (staying in ridesharing), except that, we can obtain specific travel times $tt_v(m)$ and $earning(m)$ for each matching option $m$. Example of the simulated driver matching option acceptance/rejection decisions is provided in the supplementary materials (SI Table 2). In case not all participants agree on the new matching option, it is up to the service provider to decide what to do with this matching option (e.g., driver's decision is decisive, or by majority voting).

*2.4.3. Post-matching: Order cancellation after being matched*

Driver agents are at post-matching stage if they have remaining pickup or drop-off tasks. We assume that drivers only consider order cancellation if either 1) there are some delays to their next task (either pickup or drop-off); or 2) some matched passengers cancel their orders. For simplicity, we assume that when a driver cancels a matched order, all unserved passenger requests will be canceled, and some of these requests can be assigned to the driver again in new matchings. Alternatively, one could use a secondary choice model to select the passengers to be canceled (e.g., at random, or based on time slacks). It is assumed that drivers choose the option with the highest utility.

We define the utility of driver $v_i$ canceling the existing matching option $m$ as:

$$U_{v_i,m^-} = \beta_{time,v_i} \cdot tt_v(m^-) + \beta_{cost,v_i}[cost(m^-) + CancellationFee - earning(m^-)] \quad (14)$$

where, $m^-$ indicates only traveling with onboard passengers (if any) or travelling alone. Consequently, the travel time, costs and earnings should be updated. As illustrated in Figure 2(a)-(b), we use a dynamic tree structure to store feasible routings. In this case, cancelling unserved passengers requires only traversing the dynamic tree and removing their corresponding pickup and drop-off nodes in the tree, and updating the optimal branch. In addition, a cancellation fee is introduced to discourage cancellations.

We define the utility of driver $v_i$ staying with (i.e., not cancelling) matching option $m$ in two cases. In case that the order cancellation consideration is due to delays, the utility of driver staying with existing matching option $m$ is defined as in eq. (13), except that, the delays are added to travel time $tt_v(m)$, and $cost(m)$ and $earning(m)$ are updated. In case matched passenger $r_j$ cancels the order, only pickup and drop-off node of $r_j$ will be removed from the dynamic tree, and utility eq. (13), which represents traveling without the canceled passengers, will be updated accordingly.

## 3. Ridesharing pricing bounds

In our simulation model, drivers and passengers decide to accept or reject the matching options, based on the utility of these options. We derive in this section the ridesharing pricing bounds that result in high matching accept rates.

For simplicity, we assume that the operational costs are proportional to travel times. In addition, the value of time ($VOT$) of drivers represents the combined monetary cost of travel time and operational costs. Following these assumptions, the utility of rejecting new matching options and travel alone (if not matched) $U_{v_i,0}$ for driver $v_i$ can be written as:

$$\begin{aligned} U_{v_i,0} &= \beta_{time,v_i} \cdot tt_v(v_i) + \beta_{cost,v_i} \cdot cost(v_i) \\ &= -VOT_{v_i} \cdot tt_v(v_i) \end{aligned} \quad (15)$$

And the utility of accepting matching option $m$ for $v_i$, $U_{v_i,m}$ can be written as:

$$\begin{aligned} U_{v_i,m} &= \beta_{time,v_i} \cdot tt_v(m) + \beta_{cost,v_i}[cost(m) - earning(m)] \\ &= -VOT_{v_i} \cdot tt_v(m) + earning(m) \\ &= -VOT_{v_i} \cdot tt_v(m) + p \cdot d(m) \end{aligned} \quad (16)$$

where, $p$ is the cost/km, and $d(m)$ is the billable (shared) distance of matching option $m$.

The travel time of matching option $m$, $tt_v(m)$, can be rewritten in terms of excess travel time $\Delta t(m)$ compared to traveling alone $tt_v(v_i)$:

$$tt_v(m) = tt_v(v_i) + \Delta t(m) \tag{17}$$

and the shared distance $d(m)$ can be expressed in terms of average speed $s(m)$, travel time $tt_v(m)$, and shared percentage $\alpha(m)$ of matching option $m$:

$$d(m) = \alpha(m) \cdot s(m) \cdot tt_v(m) = \alpha(m) \cdot s(m) \cdot [tt_v(v_i) + \Delta t(m)] \tag{18}$$

Therefore, the utility functions $U_{v_i,0}$ and $U_{v_i,m}$ are primarily determined by three key parameters: 1) driver value of time, $VOT_{v_i}$; 2) driver excess travel time $\Delta t(\cdot)$; 3) price, $p$. The other variables result from matching ($\alpha(m)$), and network effects ($s(m)$), which could be interdependent to these three key parameters.

Assuming that the proposed ridesharing service is not-for-profit, the price only partially compensates the driver's expenses. Under this assumption, the utilities of matching options remain negative, as follows:

$$U_{v_i,m} = \left(p \cdot \alpha(m) \cdot s(m) - VOT_{v_i}\right) \cdot [tt_v(v_i) + \Delta t(m)] \leq 0, \forall m \tag{19}$$

and then we can obtain the upper bound on the price $p$ for driver $v_i$:

$$p \leq \frac{VOT_{v_i}}{\alpha(m) \cdot s(m)}, \forall m \tag{20}$$

To have more drivers accepting matchings, a naïve approach will be finding $p$, such that the lowest utility of the matching option $\underline{m}$ is still higher than driving alone. Following the not-for-profit assumption, we have $\frac{\partial U_{v_i,m}}{\partial \Delta t(m)} = \left(p \cdot \alpha(m) \cdot s(m) - VOT_{v_i}\right) \leq 0$, therefore, the lowest utility is obtained when $\Delta t(m) = \Delta t(v_i)$, where, $\Delta t(v_i)$ is the driver $v_i$'s max excess travel time, and the lower bound can be obtained following:

$$\begin{aligned} U_{v_i,\underline{m}} &= \left(p \cdot \alpha(m) \cdot s(m) - VOT_{v_i}\right) \cdot [tt_v(v_i) + \Delta t(v_i)] \\ &\geq U_{v_i,0} = -VOT_{v_i} \cdot tt_v(v_i) \end{aligned} \tag{21}$$

and the lower bound on the price for driver $v_i$ is:

$$p \geq \frac{VOT_{v_i} \cdot \Delta t(v_i),}{\alpha(m) \cdot s(m) \cdot (tt_v(v_i) + \Delta t(v_i))} \tag{22}$$

To summarize, the pricing bounds for driver $v_i$:

$$p_{v_i} = \frac{VOT_{v_i} \cdot \Delta t(v_i),}{\alpha(m) \cdot s(m) \cdot (tt_v(v_i) + \Delta t(v_i))} \leq p \leq \frac{VOT_{v_i}}{\alpha(m) \cdot s(m)} = \overline{p_{v_i}} \qquad (23)$$

Furthermore, assuming fixed $\alpha(m)$ and $s(m)$, we can derive this lower bound $\underline{p_{v_i}}$ for driver $v_i$ with respect to the three key parameters:

$$\frac{\partial \underline{p_{v_i}}}{\partial tt_v(v_i)} < 0, \frac{\partial \underline{p_{v_i}}}{\partial \Delta t(v_i)} > 0, \frac{\partial \underline{p_{v_i}}}{\partial VOT_{v_i}} \geq 0 \qquad (24)$$

This suggests that the price lower bound $\underline{p_{v_i}}$ decreases with (travel-alone) trip durations $tt_v(v_i)$. For a relatively low $p$, drivers with longer trip durations are less likely to reject the matching options (i.e., lower matching reject rate), compared to drivers with shorter trips. Note that the price lower bound $\underline{p_{v_i}}$ increases with excess travel time $\Delta t$ and driver value of time $VOT_{v_i}$. This suggests if the price $p$ is high, drivers are more likely to accept matching options with long detours. If drivers have higher $VOT_{v_i}$, we could have them accepting matching options with higher price $p$.

Depending on the service provider's objectives, different pricing bounds could be obtained. In case the objective is to attract all drivers, the ridesharing price should be set within the following range:

$$\max_{v_i} \frac{VOT_{v_i} \cdot \Delta t(v_i),}{\alpha(m) \cdot s(m) \cdot (tt_v(v_i) + \Delta t(v_i))} \leq p \leq \min_{v_i} \frac{VOT_{v_i}}{\alpha(m) \cdot s(m)} \qquad (25)$$

By assuming the value of time for passengers quitting ridesharing as $VOT_{rv_j}$ (weighted with operational costs), and $VOT_{rv_j} > VOT_{r_j}$, following the same derivation procedure (see details in the supplementary materials), the bounds for passenger $r_j$ are defined as follows:

$$0 \leq p \leq \frac{\left(VOT_{rv_j} - VOT_{r_j}\right) \cdot tt_r(r_j) - VOT_{r_j} \cdot wt(r_j)}{\alpha(r_j) \cdot s(m) \cdot \left(tt_r(r_j) + wt(r_j)\right)} = \overline{p_{r_j}} \qquad (26)$$

where, $\alpha(r_j)$ is the average payable shared percentage based on the cost-allocation scheme (Figure 3). Moreover, we need to have non-negative passenger pricing upper bound $\overline{p_{r_j}}$:

$$\overline{p_{r_j}} = \frac{\left(VOT_{rv_j} - VOT_{r_j}\right) \cdot tt_r(r_j) - VOT_{r_j} \cdot wt(r_j)}{\alpha(r_j) \cdot s(m) \cdot \left(tt_r(r_j) + wt(r_j)\right)} \geq 0 \qquad (27)$$

, which results in an upper bound on the passenger total waiting time:

$$0 \leq wt(r_j) \leq \frac{\left(VOT_{rv_j} - VOT_{r_j}\right) \cdot tt_r(r_j)}{VOT_{r_j}} \tag{28}$$

Similar to the driver's bounds, we can derive the price upper bound $\overline{p_{r_j}}$ for passenger $r_j$ with respect to the key parameters by assuming fixed $\alpha(m)$ and $s(m)$:

$$\frac{\partial \overline{p_{r_j}}}{\partial tt_r(r_j)} > 0, \frac{\partial \overline{p_{r_j}}}{\partial VOT_{rv_j}} > 0, \frac{\partial \overline{p_{r_j}}}{\partial VOT_{r_j}} < 0, \frac{\partial \overline{p_{r_j}}}{\partial wt(r_j)} > 0 \tag{29}$$

these derivations suggest that, for a relatively high price $p$, passengers with long trip durations are more likely to accept the matching options (i.e., higher matching accept rate), since the difference in utility between traveling alone and ridesharing becomes larger. Passengers with high traveling alone $VOT_{rv_j}$ will be more tolerant to high price $p$. Conversely, if ridesharing passenger $VOT_{r_j}$ increases, the price $p$ should be lower to make ridesharing attractive. In case that $VOT_{r_j} = VOT_{rv_j}$ and $p > 0$, the utility of traveling alone is higher than ridesharing.

The relation between the waiting time $wt(r_j)$ and the price upper bound $\overline{p_{r_j}}$ can be better explained using chain rule $\frac{\partial \overline{p_{r_j}}}{\partial wt(r_j)} = \frac{\partial \overline{p_{r_j}}}{\partial VOT_{r_j}} \frac{\partial VOT_{r_j}}{\partial wt(r_j)} > 0$. The first term corresponds to low price tolerance for travelers with high ridesharing $VOT_{r_j}$. The second term implies that passengers who are willing to wait more might have lower ridesharing $VOT_{r_j}$ $\left(\frac{\partial VOT_{rv_j}}{\partial wt(r_j)} < 0\right)$.

In case the objective is to attract all passengers, the ridesharing price should be set within the following range:

$$0 \leq p \leq \min_{r_j} \frac{\left(VOT_{rv_j} - VOT_{r_j}\right) \cdot tt_r(r_j) - VOT_{r_j} \cdot wt(r_j)}{\alpha(r_j) \cdot s(m) \cdot \left(tt_r(r_j) + wt(r_j)\right)} \tag{30}$$

Finally, to ensure pricing feasibility, the smallest passenger price upper bound should be larger than the largest driver price lower bound:

$$\max_{v_i} \underline{p_{v_i}} \leq p \leq \min_{r_j} \overline{p_{r_j}} \tag{31}$$

We illustrates the pricing bounds with a toy network in the supplementary materials.

## 4. Numerical Experiments

In this section, we perform numerical experiments to illustrate the capabilities of our proposed ridesharing simulation model. The following subsections describe respectively, the experimental setup, driver agent activities, simulation results for the default setup, and sensitivity analysis with respect to pricing and matching time window.

### *4.1. Experiment setup*

The experiments are conducted using the well-known Winnipeg network, which consists of 154 zones, 1067 nodes, 2535 links, and 4345 origin-destination (OD) pairs. The total hourly demand of the Winnipeg matrix is 54,459 trips (Bekhor et al. 2008), the OD spatial distribution and trip distance distribution are shown in Figure 8. It is also assumed that the hourly demand is distributed into four 15-minute periods, with each period accounts for 10%, 40%, 40% and 10% of the total hourly demand, respectively. We assume the hourly demand starts from 8:00 and time-dependent travel times associated with this demand pattern are also imported to the simulation. All the experiments are run on a 6-core PC with 32 GB ram, with the simulation model written in Python.

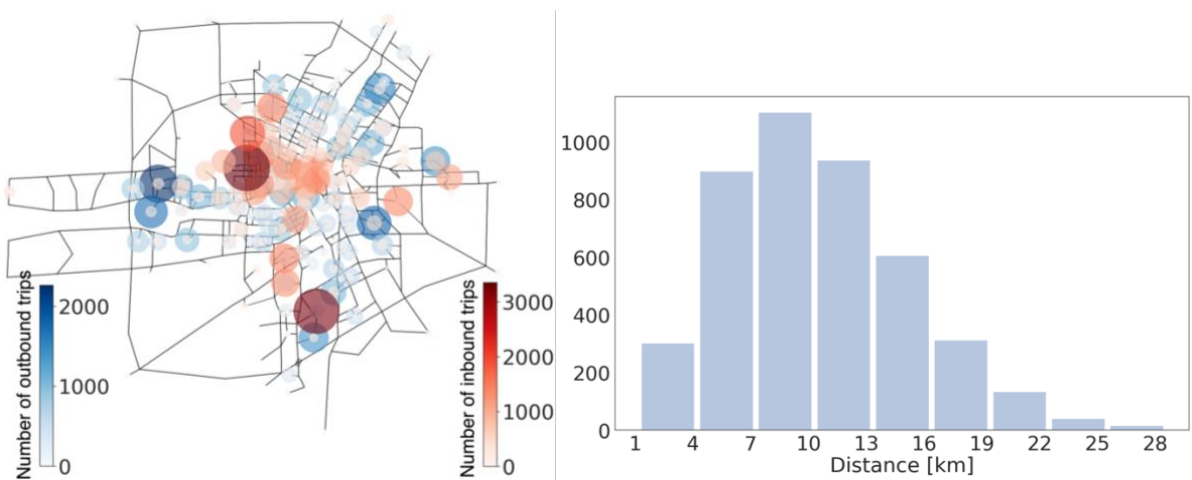

Figure 8 Distribution of OD matrix trips and distances

To obtain insights of the simulation process, we assume a default setting with 2,000 ridesharing passenger trips and 1,600 ridesharing driver trips. These trips are randomly sampled from the 54,459 trips without replacement, and distributed into four time periods by 10%, 40%, 40% and 10%. For simplicity, we assume agents departs uniformly within each period, and their main attributes are specified as in Table 2.

Table 2 Agent attributes

| Agent main attribute | Attribute value |
|---|---|
| Value of time [$\beta_{time}/\beta_{cost}$] | $VOT_{r_j} \sim Uniform(0.5, 1)$ |
| | $VOT_{v_i} \sim Uniform(3, 3.5)$ |
| Passenger perceived travel alone unit cost [cost/time] | $cost'_{travel\ alone} \sim Uniform(3, 3.5)$ *such that,* $\left( cost(r_j) = tt_r(r_j) \times cost'_{travel\ alone} \right)$ |
| Passenger perceived ridesharing unit cost $\left[ \beta_{\exp\ pay, r_j} \cdot cost'_{travel\ alone} \right]$ | $cost'_{ride-sharing} \sim Uniform(1.5, 3.5)$ *such that,* $\left( \beta_{\exp\ pay, r_j} \cdot cost(r_j) = tt_p(r_j) \times cost'_{ride-sharing} \right)$ |
| Max excess travel time $\Delta t(v_i)$ | *Solve* $U_{v_i,0} = U_{v_i,1}$ *with* $\beta_{\exp\ earn, v_i} = 1$, *for an upper bound* $\max \Delta t(v_i)'$, *and set* $\max \Delta t(v_i) \sim \min(Uniform(15, 45), \max \Delta t(v_i)')$ |
| Max total waiting time $wt(r_j)$ *(pickup and matching waiting times)* | *Solve* $U_{r_j,0} = U_{r_j,1}$ *for an upper bound* $\max wt(r_j)'$, *and set* $\max wt(r_j) = \min(Uniform(5, 10), \max wt(r_j)')$ |

As discussed in section 2.4.1, drivers (deterministically) choose to quit ridesharing if this option has the highest utility. Therefore, the upper bound on excess travel time $\Delta t(v_i)'$ can be found by solving $U_{v_i,0} = U_{v_i,1}$, after which the utility of quitting the service will be larger than staying in it. The driver's individual-specific max $\Delta t(v_i)$ is simulated and bounded by $\Delta t(v_i)'$. Similarly, by solving $U_{r_j,0} = U_{r_j,1}$, we can obtain the upper bound for the passenger's total waiting time $wt(r_j)'$, and simulate the passenger's individual-specific max $wt(r_j)$ within this bound.

Furthermore, we assume a default unit cost of a ridesharing trip to be 6 [unit cost/km]. The consideration behind this assumption is to compensate the extra detours for ridesharing drivers, and increase the matching option accept rate. Assuming an average link travel time of 1.5 [min/km] with the assumed unit price, the resulting value of time for ridesharing (4 [cost/min]) will be higher than the driver's travel alone value of time (3-3.5 [cost/min] as in Table 2).

## *4.2. Driver agent activities*

In this subsection, a driver agent from the experiment is selected to illustrate the behavior of our ride-sharing simulation model (Figure 9). In this example, there is a single driver and 5 possible ridesharing passengers. The matching option acceptance/rejection and order cancellation behaviors are also showcased in this example.

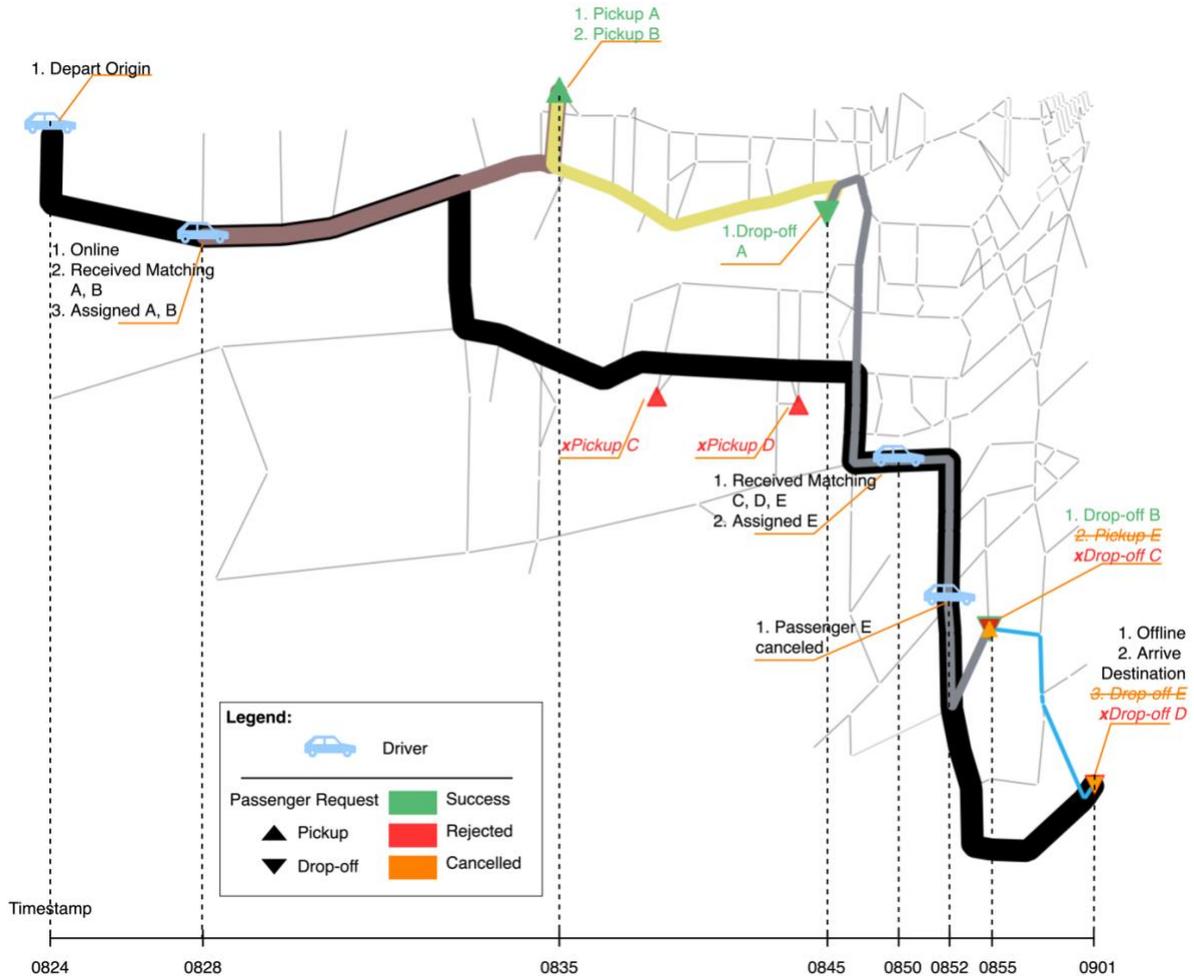

Figure 9 Description of a driver Agent

The driver agent $v_1$ with capacity 4 (passengers) departs his/her destination at 08:24, with his preferred itinerary (marked in black). At 08:28, the driver agent sends a ridesharing request, and immediately receive a matching option $(v_i, r_A, r_B)$ to take passengers A and B (together). The driver and both passengers A and B accept the matching option. The driver then follows the guided route (marked in brown) to pick up passengers A and B. Passengers A and B are successfully picked up at 08:35. The driver then follows the new route (marked in yellow) to drop off passenger A at 08:45. After that, the driver has one onboard passenger B, and travel on the grey route to drop off passenger B.

At 08:50, the driver receives a new matching option $(v_1, r_B, r_C, r_D, r_E)$, which is associated with the onboard passenger B, and additional new passengers C, D and E. Meanwhile, passengers B, C, D, E also receive this matching option for their acceptance/rejection decisions. The service provider receives their responses as in Table 3.

Table 3 Example of agent decisions on a matching option

| Matching option $m$ | $v_1$ | $r_B$ | $r_C$ | $r_D$ | $r_E$ |
| --- | --- | --- | --- | --- | --- |
| $(v_1, r_B, r_C, r_D, r_E)$ | Accept | Accept | Reject | Reject | Accept |

Since not all new passengers accept the matching option, the service provider modifies the matching option to be $(v_1, r_B, r_E)$, and updates driver's routing (see Supplementary materials). The pickup of new passenger E is scheduled at 08:52 and requires no extra detour, so the driver continues traveling on the grey route.

However, due to delays on pickup (i.e., driver did not arrive at 08:52), passenger E decides to cancel the matched order. The remaining participants of the matching (driver $v_1$ and passenger B) receive notifications about the cancellation and are asked for their decisions to stay/cancel the matching. Since all onboard passengers should be delivered, the driver continues traveling the grey route to drop off passenger B at 08:55 and reaches his/her own destination at 09:01 using the blue route.

The above example illustrates how the driver agent is progressed in the simulation model, and how the service provider dynamically matches drivers and passengers. This example also accounts for matching option acceptance/rejection, and order cancellation behaviors. In the following subsection, we present the experiment results at the network level.

*4.3. Simulation results*

In this subsection, we evaluate the simulation using 11 performance metrics, among which option acceptance/rejection (driver/passenger new matching option accept rate), and order cancellation (Matching execution rate) are considered. We perform 10 replications for the default simulation setup (as in Section 4.1), and the mean and standard deviation of the performance metrics are reported in Table 4.

Table 4 Simulation evaluation (averaged over 10 replications)

| Performance attribute | Mean | Standard deviation |
|---|---|---|
| **Simulation runtime [min]** | 18.1 | 0.4 |
| **Driver percentage excess travel time** $$\frac{1}{|V|}\sum_{v_i}\frac{at(v_i) - tt(v_i)}{tt(v_i)} \quad (32)$$ where, $at(\cdot)$ is the agent's actual travel time from his/her origin to the destination, and $tt(\cdot)$ is the agent's shortest travel time. | 19.1% | 0.3% |
| **Passenger percentage excess travel time** $$\frac{1}{|R|}\sum_{r_j}\frac{at(r_j) - tt(r_j)}{tt(r_j)} \quad (33)$$ | 21.2% | 2.6% |
| **Driver success rate** $$\frac{\sum_{v_i}\gamma(v_i)}{|V|} \quad (34)$$ $\gamma(v_i) = 1$, if driver $v_i$ delivered at least 1 passenger, and $\gamma(v_i) = 0$ otherwise. | 41.8% | 0.7% |
| **Passenger success rate** $$\frac{\sum_{r_j}\gamma(r_j)}{|R|} \quad (35)$$ $\gamma(r_j) = 1$, if passenger $r_j$ reaches destination using ridesharing, and $\gamma(r_j) = 0$ otherwise. | 44.5% | 0.7% |
| **Overall success rate** $$\frac{\sum_{v_i}\gamma(v_i) + \sum_{r_j}\gamma(r_j)}{|V| + |R|} \quad (36)$$ | 43.3% | 0.7% |
| **Driver new matching option accept rate** $$\frac{1}{|V|}\sum_{v_i}\frac{\sum_k \pi_k(v_i) \cdot \lambda_k(v_i)}{\sum_k \lambda_k(v_i)} \quad (37)$$ where, $\lambda_k(\cdot) = 1$, if agent is matched at $k$th matching, and $\lambda_k(\cdot) = 0$ otherwise; and $\pi_k(\cdot) = 1$ if agent accepts any matching option at $k$th matching, and $\pi_k(\cdot) = 0$ otherwise. | 45.7% | 1.3% |
| **Passenger new matching option accept rate** | 51.6% | 1.0% |

$$\frac{1}{|R|}\sum_{r_j}\frac{\sum_k \pi_k(r_j)\cdot \lambda_k(r_j)}{\sum_k \lambda_k(r_j)} \qquad (38)$$

**Average matching rate**

Average percentage of agents being matched by the service provider:

$$\frac{1}{|K|}\sum_{k\in K}\frac{\sum_{v_i}\lambda_k(v_i)+\sum_{r_j}\lambda_k(r_j)}{|V|_k+|R|_k} \qquad (39) \qquad 11.3\% \qquad 0.6\%$$

where, $K$ is the total number of matchings, $|V|_k$ and $|R|_k$ record the number of drivers and passengers in the matching pool at $k$th matching.

**Overall Matching rate**

$$\frac{\sum_{v_i}\lambda(v_i)+\sum_{r_j}\lambda(r_j)}{|V|+|R|} \qquad (40) \qquad 66.5\% \qquad 0.7\%$$

where, $\lambda(\cdot)=1$, if $\sum_k \lambda_k(\cdot)\geq 1$, and $\lambda(\cdot)=0$ otherwise.

**Matching execution rate**

$$\frac{1}{|V|}\sum_{v_i}\frac{\sum_{r_j\in R_{v_i}}\gamma(r_j)}{|R_{v_i}|} \qquad (41) \qquad 59.5\% \qquad 1.4\%$$

where, $R_{v_i}$ is the set of passengers being matched with driver $v_i$, and $\sum_{r_j\in R_{v_i}}\gamma(r_j)$ represents the number of passengers being delivered by driver $v_i$.

**Percentage VKT saving**

$$\frac{\sum_{v_i}[d(v_i)-sd(v_i)]-\sum_{r_j}\gamma(r_j)sd(r_j)}{\sum_{v_i}sd(v_i)+\sum_{r_j}sd(r_j)} \qquad (42) \qquad 7.0\% \qquad 0.4\%$$

where, $d(v_i)$ represents the actual VKT of driver $v_i$ (for either ridesharing or drive alone), and $sd(\cdot)$ represents shortest distance for traveling alone.

For simulating the specified driver supply and passenger demand, the average simulation runtime is about 18 minutes for simulating 1.5-hour travel demand. In the simulation, 45 ridesharing matchings (every 2 minutes in the simulation) are performed on average. The above results suggest that our simulation model has the potential for ridesharing operations.

The passengers' success rate (44.5%) is higher than drivers' success rate (41.8%) on average. This could be related to passengers' higher flexibilities in detours for reducing their monetary costs. This is also indicated by the relatively higher passenger new matching option accept rate (51.6%) than the drivers' (45.7%), suggesting passengers are more tolerant to lower quality matchings (e.g., relatively large percentage excess travel time, 21.2%) than the driver (19.1%).

The average matching rate (11.3%) shows how many ridesharing drivers and passengers the service providers could match, and these matching options are accepted in each matching run. This value is relatively low, compared to Yao and Bekhor (2021), and can be explained by the newly introduced dynamic supply/demand capability in the simulation. There is a significant difference between the overall matching rate (66.5%) and the overall success rate (43.3%), suggesting that the dynamic and driver-passenger actions should also be considered for developing an effective ridesharing system.

The matching execution rate (59.5%) indicates how many matchings are executed and not cancelled. Note that, the relatively low VKT savings (7% for a 43.3% overall success rate) is due to the fact that drivers make relatively long detours (19.1% in the experiment) to pick up and drop off passengers, and they may make extra detours without any passenger due to order cancellations (as suggested in Table 4, only 59.5% of the matchings are executed).

Our experiment gives rich insights to the ridesharing operation and provides new opportunity and direction for improving the ridesharing systems. Additional simulation results of driver agents, passenger agents, and service providers are provided in supplementary materials. In the following subsection, we analyze key ridesharing factors.

### 4.4. Sensitivity Analysis

This section performs a sensitivity analysis with respect to selected factors that affect agent matching option acceptance/rejection and order cancellation behaviors, and consequently the ridesharing system performances (Table 5). For simplicity, we discuss two main factors: 1) pricing, which affects agent matching option acceptance/rejection; 2) ridesharing matching window, which affects order cancellations. To account for different OD distributions, 10 replications are conducted for all the tests, in which ridesharing drivers and passengers are sampled. Each point in the following figures represents the average value of 10 replications and the shades indicate standard deviations. Additional sensitivity analysis on driver supply is also provided in supplementary materials.

Table 5 Selected factors and their levels for the sensitivity analysis

| Factor | Default Level | Analysis Levels |
|---|---|---|
| Pricing | 6 [cost/km] | 0.5, 1, 2, 4, 6, 8, 10, 12, 14 [cost/km] |

| Ridesharing matching window | 2 minutes | 0.1, 0.5, 1, 2, 3, 4, 5, 6, 7, 8 minutes |

### *4.4.1. Impact of pricing and its bounds*

The impact of the ridesharing prices and the pricing bounds for which drivers and passengers are likely to accept the matching options are analyzed. As shown in eq. (23) and (26), the drivers and passenger price bounds are determined by a) simulation parameters: value of times $VOT_{(\cdot)}$; driver excess travel time $\Delta t(\cdot)$; passenger total waiting time $wt(\cdot)$; and price $p$; b) interdependent variables of matching ($\alpha(m)$), and network effects ($s(m)$). To evaluate the impact of pricing, we purposely fix the other simulation parameters, and vary only the unit price $p$ (which consequently also affects matchings). Furthermore, we segment the ridesharing demands by travel time intervals of 5 min (with uniform sampling of each demand group), to demonstrate the effect of trip times on the pricing bounds (eq. 24 and 29). We show in Figure 10a the impact of pricing on driver matching option accept rate for each demand group (by trip time durations), Figure 10b illustrating the rejection rate at unit price of 0.5.

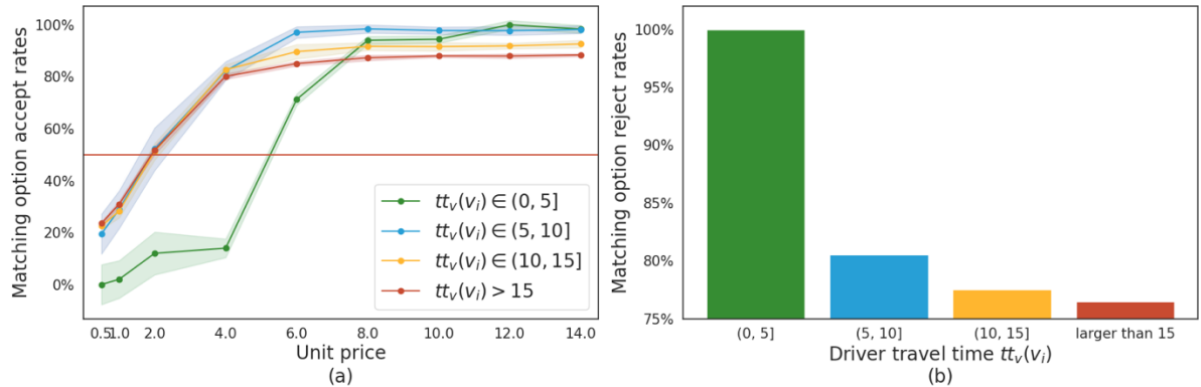

Figure 10 Impact of pricing on driver matching acceptances

For all driver demand groups, the matching option accept rates increase with the unit price (Figure 10a), which is due to the increase of ridesharing utility eq. (13). Moreover, as derived in eq. (24), when price $p$ is low (e.g., 0.5 cost/km in this experiment), drivers with longer trip durations are less likely to reject the matching options (Figure 10b).

By the assumed binary accept/reject setup, the price lower bound $p_{v_i}$ can be obtained for a specific driver $v_i$ when the acceptance probability is 50%. The matching option accept rate (eq. 20) can be interpreted as the aggregated acceptance measure over all the driver agents. In this case, the overall price lower bound can be approximated with prices that result in 50% matching option accept rate (red line in Figure 10). Therefore, the price lower bound for drivers with very short trips ($tt_v(v_i) \leq 5$) is about 5.5 cost/km. The price lower bound for drivers with relatively longer trips ($tt_v(v_i) > 5$) is only about 2 cost/km. This suggests that a relatively high price is needed to attract drivers with shorter trip durations to participate in ridesharing. Similarly, we show the impact of pricing on passenger matching option accept rate for each demand group in Figure 11. Note that, this figure shows combined effects of various components influencing the pricing bounds, including the key parameters and interdependent matching and network effect variables.

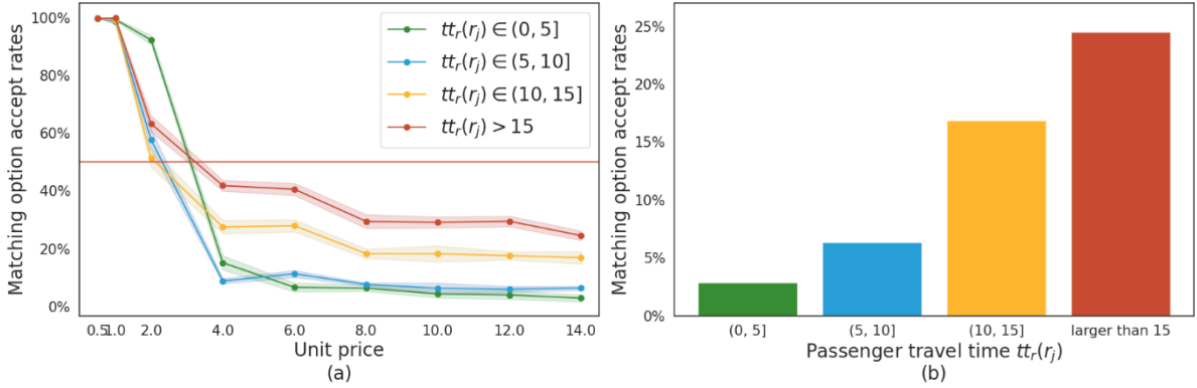

Figure 11 Impact of pricing on passenger matching acceptances

Results indicate that the matching option accept rates decrease with the unit price for all demand groups (Figure 11a). In Figure 11b, in accordance with the derivation of eq. (29), for a relatively high price ($p = 14$ in this experiment), passengers with long trip durations are more likely to accept the matching options. By approximating the overall price upper bound with prices that result in 50% matching option accept rate, we observe that passengers with very long ($tt_r(r_j) > 15$) and very short trip durations ($tt_r(r_j) \leq 5$) have a price upper bound about 3.5 cost/km. passengers with medium trip durations ($5 \leq tt_r(r_j) < 15$) have a price upper bound about 2.5 cost/km.

Note that in Section 3 eq. (29), the upper bound of passenger price is positively correlated to their trip durations $\left(\frac{\partial \overline{p_{r_j}}}{\partial tt_r(r_j)} > 0\right)$. This is obtained by assuming a fixed $\alpha(r_j)$. The results in Figure 11 suggest a complex interdependency between $\alpha(r_j)$ and other parameters, such as, $tt_r(r_j)$. In this case, using chain rule $\frac{\partial \overline{p_{r_j}}}{\partial tt_r(r_j)} = \frac{\partial \overline{p_{r_j}}}{\partial \alpha(r_j)} \frac{\partial \alpha(r_j)}{\partial tt_r(r_j)}$, we discuss these trade-offs. The first term captures the effect of payable portion $\alpha(r_j)$ on the pricing upper bound. Using eq. (26), $\frac{\partial \overline{p_{r_j}}}{\partial \alpha(r_j)} \leq 0$ is obtained, which suggests passengers with very short trip durations could have relatively low payable portions $\alpha(r_j)$. The second term captures the trade-off between payable portion $\alpha(r_j)$ and trip duration $tt_r(r_j)$. To understand this trade-off, the passenger average matching rates are shown for each passenger demand group (Figure 12).

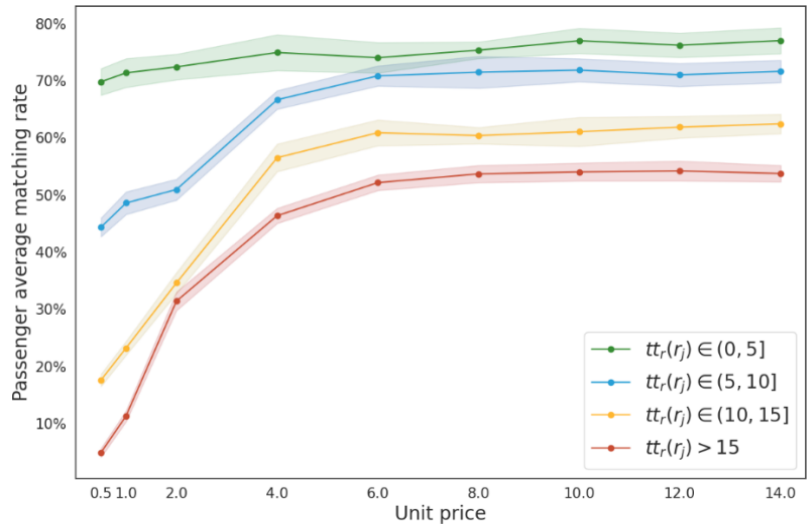

Figure 12 Effects of unit price for passenger success rate in each demand group

We observe in Figure 12 that passengers with long trip durations have low average matching rate, regardless of the unit price, which consequently could result in higher payable portions $\alpha(r_j)$ (i.e., the unit cost split between less passengers). In this case, we can approximate this relation as $\frac{\partial \alpha(r_j)}{\partial tt_r(r_j)} \geq 0$. These two terms represent the trade-offs between payable portions $\alpha(r_j)$ and trip duration $tt_r(r_j)$, with different signs in their derivatives.

*4.4.2. Impact of ridesharing matching window*

The impact of the duration of matching windows on the ridesharing system performances is analyzed. Figure 13 presents six graphs with selected metrics: (a) driver/passenger percentage excess travel time; (b) driver, passenger and overall success rate; (c) passenger expected pickup waiting time; (d) driver/passenger new matching option accept rate; (e) matching execution rate, overall matching rate, and overall matching rate; (f) percentage VKT savings and number of passengers leaving before matching. Several insights can be observed from these results and are discussed in the following.

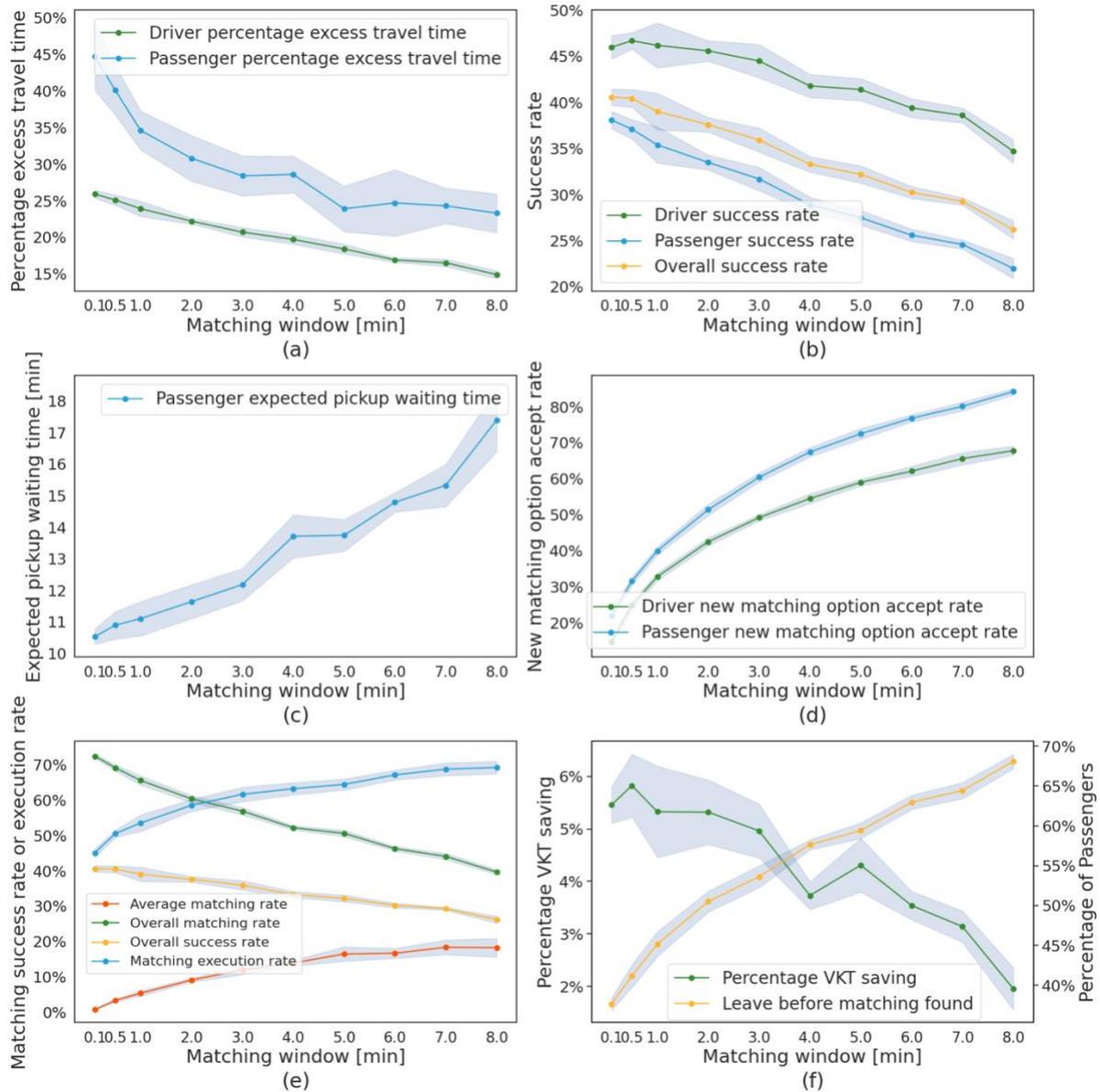

Figure 13 Performance metrics for ridesharing matching window sensitivity analysis

Figure 13(a, d, e) show the positive effects on increasing matching time window. With increases in time window, we are expecting to have higher quality matchings (i.e. shorter detours for the matched drivers and passengers, as in Figure 13(a)), therefore have higher matching option accept rates (Figure 13(d)), and higher matching execution rates (Figure 13(e)). The higher matching execution rates also indicate smaller gaps between the overall matching rate and overall success rate (Figure 13(e)), suggesting less overestimates in system performances.

The overall matching success rate also gradually increases with larger matching windows (Figure 13 (e)). This result is consistent with the empirical matching probability in the literature, in which, the matching probability of ridesharing is modeled as $p(q_r, \phi) = 1 - \exp(-\eta q_r \phi)$, with $\eta$ as the scale parameter, $q_r$ as ridesharing demand rate, and $\phi$ as the matching window (Yan et al., 2020; Ke et al., 2021).

However, these positive effects come at the expense of longer matching option waiting times (i.e., the drivers and passengers sent their ridesharing requests and were told to wait for matching). The longer waiting times may result drivers and passengers leave the ridesharing system before the service provider sends them any matching option (Figure 13(f)), and decreases the driver, passenger, and overall success rate (Figure 13(b)), which in turn causes the average expected waiting time to be longer (Figure 13(c)).

The negative effect of longer matching window on percentage VKT savings (Figure 13(f)) could be counterintuitive at first glance. One may expect that, with longer matching time, more drivers and passengers could be considered in the matching, and result higher quality matchings, i.e. shorter detours and increase of VKT savings (the matching objective function). However, this is only correct if the passengers are willing to wait for a relatively long time before he/she receives a matching option, i.e., passenger order cancellation is not considered. As opposed to Yan et al., 2020 and Ke et al., 2021, our model endogenously captures ridesharing demand as a function of matching window $q_r(\phi)$ as well. When the matching window increases, the overall matching rate increases for the drivers and passengers remaining in the matching pool. However, in this case, more drivers and passengers left the system before matching, which results in less potential shared trips, and consequently lower overall success rate and percentage VKT savings.

We illustrate the potential application of the proposed ridesharing simulation model for designing and evaluating operational strategies, in the case of sensitivity analysis. The results emphasize that unrealistic assumptions related to driver and passenger behavior could lead to misleading conclusions overestimating the performance of a ridesharing system. For example, if ignoring cancellations, one would choose a larger matching window based on metrics, for example, like the overall matching rate. However, larger matching window could cause drivers and passengers leaving services even before matching, and result in poor "real-world" performance (e.g., lower overall success rate). Our simulation model provides a realistic and comprehensive testbed for ridesharing service developments.

## 5. Comparison with previous approaches

In this section, we illustrate the importance of considering dynamic supply-demand interactions in designing and evaluating ridesharing systems, by explicitly comparing the performance metrics with three models: 1) static model (as in Yao and Bekhor, 2021); 2) dynamic model without interactions (e.g., Herbawi et al., 2012; Alonso-Mora et al., 2017; Masoud et al., 2017; Simonetto et al., 2019); 3) dynamic model with interactions (this paper).

Table 6 Model setups for the comparison experiment

| Attributes | Static model | Dynamic without interaction | Dynamic with interaction |
| --- | --- | --- | --- |
| Time-dependent Demand | *Not relevant* | √ | √ |
| Pre-matching cancellation | *Not relevant* | *Not relevant* | √ |
| Matching accept/reject | *Not relevant* | *Not relevant* | √ |
| Post-matching cancellation | *Not relevant* | *Not relevant* | √ |
| Demand | 100 passengers | 10, 40, 40, 10 passengers | 10, 40, 40, 10 passengers |
|  | 50 drivers | 5, 20, 20, 5 drivers | 5, 20, 20, 5 drivers |
| Trip time [min] | 28.5 | 28.5 | 28.5 |
| Shortest distance [km] | 34.6 | 34.6 | 34.6 |
| $\Delta t(v_i)$ | $Uniform(15, 45)$ | $Uniform(15, 45)$ | $Uniform(15, 45)$ |
| $wt(p_j)$ | $Uniform(5, 10)$ | $Uniform(5, 10)$ | $Uniform(5, 10)$ |
| Capacity | Sampling [1, 2, 3, 4] | Sampling [1, 2, 3, 4] | Sampling [1, 2, 3, 4] |

We randomly select one OD pair of the Winnipeg network and generate the same set of driver and passenger agents for the 3 models. We assume 100 passengers and 50 drivers for this OD pair, which are distributed into four time periods by 10%, 40%, 40% and 10% in the dynamic models (Table 6). Moreover, for this example we postulate that drivers and passengers have only 60% chance to accept the matching, which is used to compute a predefined "true" ridesharing success rate. That is, regardless the utility of matching option $m$, we purposely assume independent drivers with $P_{v_i}(m) = 60\%$ and passengers with $P_{r_j}(m) = 100\%$, which results in a joint probability for matching option acceptance $P(m) = P_{v_i}(m) \prod_{r_j \in m} P_{r_j}(m) = 60\%$. Given a hypothetical case that all drivers and passengers are matched together in one matching (e.g., the static case), the overall matching rate = 100%. Assuming average occupancy of 2 passengers, the matching option accept rate is 60% (by aggregating the acceptance probability over the matching options), this means 30 drivers (= 60%accept × 100%matched × 50drivers), and 60 passengers (30drivers × 2passengers/driver) finish their trips with ridesharing. Therefore, the success rate will be 60%.

To simulate different ridesharing agents, the spatiotemporal and capacity constraints are sampled. We perform 10 runs for each model and compute the average and standard deviation results. The matching rate and simulated rates are shown in Table 7, and the predefined "true" success rate is defined as 58% (= 60% × matching rate of the static model, 96.67%, which is the maximum matching rate considering the spatiotemporal constraints).

Table 7 Comparison of model simulated success rates and predefined "true" success rate

| | Static | | Dynamic without interaction | | Dynamic with interaction | |
|---|---|---|---|---|---|---|
| Metrics | Mean | Stdev | Mean | Stdev | Mean | Stdev |
| Overall matching rate | 96.67% | 0.64% | 90.11% | 1.98% | 86.72% | 2.70% |
| Simulated success rate | - | - | 90.11% | 1.98% | 60.39% | 5.21% |
| t-test(58%) *predefined "true" success rate* 58% = 60% × 96.67% | 60.42 | | 16.22 | | 0.46 | |

The results show that, the proposed dynamic model with interaction, could replicate the predefined "true" success rate (i.e., the simulated success rate is not statistically significantly different from 58% with critical value of 1.96), while other models overestimate the ridesharing system performance.

For the single OD pair example, we compare vehicle trips, VKT and vehicle hour traveled (VHT) for the base case (Winnipeg network) and the three models in Table 8. We assume in the base case that all travelers travel alone (150 vehicle trips) on their shortest route, while ridesharing is considered in the other three models.

Table 8 Model comparison in vehicle trips, VKT, and VHT (single OD pair)

| Metrics | Base case  *No ridesharing* | Static | | Dynamic without interaction | | Dynamic with interaction | |
|---|---|---|---|---|---|---|---|
| | | Value | Saving | Value | Saving | Value | Saving |
| Vehicle Trips | 150 | 52 | 65.3% | 53 | 64.7% | 63 | 58.0% |
| VKT | 5190 | 1796 | 65.4% | 2157 | 58.4% | 2329 | 55.1% |
| VHT | 71.25 | 24.67 | 65.4% | 31.07 | 56.4% | 32.93 | 53.8% |

Results show that, static model has the highest savings in vehicle trips, VKT and VHT, compared to the two dynamic models, because all driver and passenger requests are considered in the static matching. In the dynamic case, model with driver-passenger interactions has relatively lower savings in vehicle trip, VKT and VHT, which corresponds to the relatively lower ridesharing success rate. This relatively less savings suggests potential overestimates of ridesharing network benefits if the driver-passenger interactions are not properly captured.

## 6. Summary and Outlook

This paper presents a new ridesharing simulation model that accounts for dynamic driver supply and passenger demand, and complex interactions between drivers and passengers. Specifically, this paper is one of first models that explicitly considers driver and passenger acceptance/rejection on the matching options, and cancellation before/after being matched. New simulation events, procedures and modules have been developed to handle these realistic interactions in this paper.

The capabilities of the proposed ridesharing simulation model are illustrated using numerical experiments in the well-known Winnipeg network. The detailed ridesharing simulation progress is demonstrated, with examples of agent matching option acceptances/rejections and order cancellations action behaviors.

Ridesharing pricing bounds are analytically derived based on the assumptions of driver and passenger matching option acceptance/rejection behaviors and exemplified in a toy network.

The proposed simulation model, which explicitly considers agent matching option acceptance/rejection behaviors, is then used to evaluate pricing bounds for real-size problems. Results show that, high price is required to incentive drivers with short trip durations to participate in ridesharing, and passengers with rather long or short trips could accept higher price.

Further sensitivity analysis on matching time window exemplifies the importance to capture agent order cancellation behaviors. For example, average matching rate and execution rate increases as matching window increases for drivers and passengers remaining in ridesharing matching pool. Under our experiment setting, larger matching window will also cause drivers and passengers leaving ridesharing even before the matching occurred, and consequently result in lower overall success rate and lower VKT savings.

Moreover, success rates simulated using different models are compared with a predefined "true" success rate. Results show that, only the proposed simulation model, with dynamic supply-demand interactions, could replicate the predefined "true" success rate, while other models overestimate the performance of a ridesharing system. These results illustrate the need to consider dynamic supply-demand interactions, which are often overlooked and have strong impacts (see Alonso-González et al., 2020). Without these interactions, a realistic ridesharing system performance cannot be evaluated properly, which could cause misconfiguration when designing a ridesharing system (e.g., matching window as shown in the previous section), and under-utilize the potentials of ridesharing. Our simulation model provides such insights for developing more efficient and effective operational strategies.

The numerical experiments illustrate the diverse functionalities that the proposed ridesharing simulation model can provide and suggest a wide range of potential applications. For example, evaluating the effectiveness of the ride-sharing matching algorithm when driver and passenger acceptance/rejection are considered, designing personalized ridesharing trip planning, dynamic pricing scheme, and matching window with considerations on driver/passenger cancellations.

The simulation model could be extended to allow more service options, such as, ride-hailing, ride-splitting, and even autonomous mobility-on-demand. These new driver agents could be created by relaxing some ridesharing constraints. This extension would allow the analysis of the share of different service types in a more comprehensive setting. In the current version of the simulation model, only one service provider agent was created. The simulation can handle additional service providers and allows studying the competitions among them.

The proposed simulation model considers the case that, service providers need not to optimize the drivers' and passengers' utilities. In essence, matching option rejection occurs when the objective of the service provider does not align with the drivers' and passengers' objective (utility maximization). In the case that agent utility maximization is set as service provider's objective, the ridesharing matching problem to a certain extend resembles the stable matching problem (Wang et al., 2018), in which agents cannot improve their utilities by switching matching options, and consequently no rejection occurs. Future research can investigate the effects of different matching objectives on the matching acceptance rate.

Recently, there have been a heated discussion on the impact of shared mobility on network congestion (Wei et al., 2020; Ke et al., 2020b; Sun and Szeto, 2021), and on the impact of driver and passenger's pre-day mode choice (Djavadian and Chow, 2017a, b; Liu et al., 2019; Nahmias-Biran et al., 2020) on ridesharing system performance. Future research will incorporate traffic dynamics into the simulation model and consider agent pre-day and within-day mode choice behaviors.

# A ridesharing simulation platform that considers dynamic supply-demand interactions
— Supplementary materials —

## 1. RIDESHARING SIMULATION PLATFORM

### 1.1. Object framework

The proposed ridesharing simulation platform follows the agent-based modeling approach, as shown in SI Figure 1. The object-oriented framework includes seven types of classes: **Passenger**, **Stop**, **Trip**, **Itinerary**, **Service Provider**, **Driver**, and **Schedule** as described below.

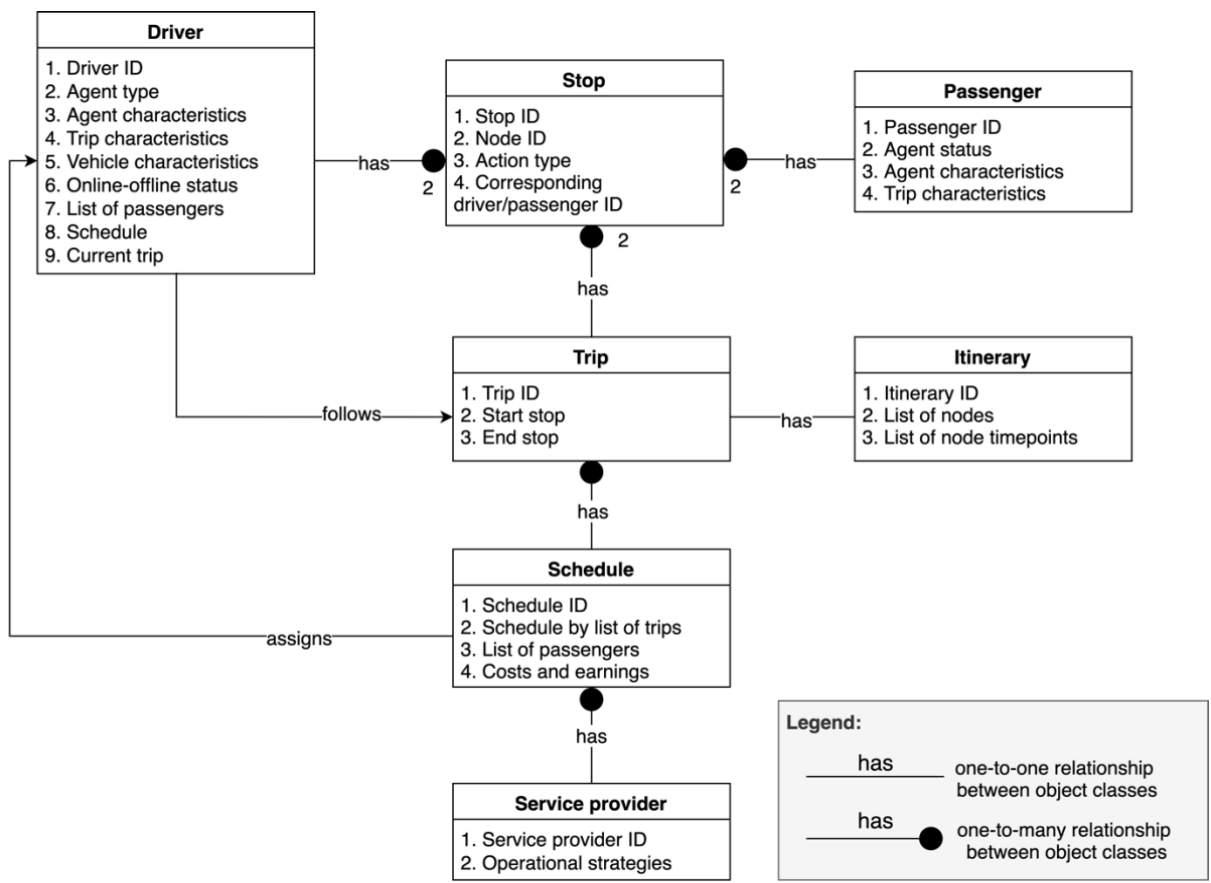

SI Figure 1 Object-oriented framework for the ridesharing simulation platform

The Passenger agents are instances of the **Passenger** class, defined as follows:

- Agent status: the Passenger has four status – *not matched*, *matched waiting for pickup*, *on-board, dropped-off*.
- Agent characteristics: sociodemographic attributes and personal constraints on ridesharing service. For example, the agent value of time, earliest/latest departure time, maximum riding time (or distance), and latest arrival time.
- Trip characteristics: it includes the origin, destination, current location, departure time, requested number of seats.

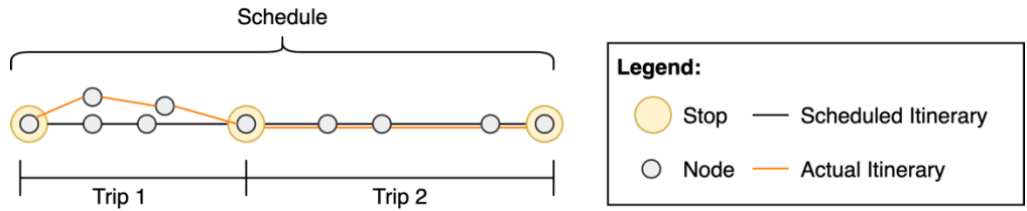

SI Figure 2 Example of Stop, Trip, Itinerary and Schedule

An example of relations between Stop, Trip, Itinerary and Schedule is shown in SI Figure 2. Each Passenger and Driver are associated with two **Stop**s, which defines the locations of their origin and destination. Each **Trip** is composed of the starting Stop and the ending Stop. Given a Trip, two **Itineraries**, which hold a list of nodes and the corresponding timestamps, are generated for such Trip. One of the two **Itineraries** represents the scheduled **Itinerary** with average/estimated network travel time. And the other one represents the actual **Itinerary** modified with agent-specific driving speed/route preference. These **Itineraries** provide additional flexibility in modeling driver heterogeneity and can be used for updating travel times. A **Schedule** is a combination of Trips that determines the Driver's pickup and drop-off sequence, and its corresponding Passenger-cost and Driver-earning information.

The **Driver** class defines the compositions of a driver agent by:

- Agent type: drive alone Driver or ridesharing Driver. Drive alone drivers do not participate in ridesharing services and travel directly to their destinations. Whereas ridesharing Drivers are willing to take peer Passengers with them.
- Agent characteristics: like the Passenger class, Driver class agent characteristics defines the agent value of time, maximum total number of passengers willing to take, maximum driving time (or distance), latest arrival time, etc.
- Trip characteristics: like the Passenger class, it includes the origin, destination, current location, departure time.
- Vehicle characteristics: attributes related to the vehicle, e.g., vehicle capacity.
- Online-offline status: defines whether the Driver is accepting new matchings or not. Together with the agent characteristics, the agent may choose to be *online* to receive new matching options; or to be *offline*, which means they decide to only finish the assigned matchings and do not want to receive new matching options.
- List of Passengers: defines the Passengers that are matched with the Driver and can be used to interact with Passengers in case of delays, changes in matching, or order cancellations.
- Schedule: It holds information of a sequence of pickup and drop-off Stops the Driver should visit for his/her matched Passengers, a list of Passengers, and the cost/earning information.
- Current Trip: it holds information about which pickup/drop-off Stop the Driver is heading to, and it is also associated with an Itinerary.

During the simulation, the Driver agent will update the Itinerary timestamps of current Trip, and update matching information to all relevant Passengers. Once the Trip is completed, the Driver agent will update its capacity and update the corresponding Passenger agent to be on-board or dropped-off. In case of cancellation (either by driver or passenger), the Driver agent will also determine the new Schedule and compute new itinerary for the updated current trip. With these updated Schedule and matching information, Driver and Passenger decide whether to cancel the order, or even trigger no-show.

The last agent is the Service Providers, which may have different operational strategies, for example, different matching algorithms, pricing schemes and so on. Based on **Service Provider**'s operational strategies, and the Passenger/Driver ridesharing matching request inputs, the Service Provider matches Drivers and Passenger, and generates **Schedule** as part of the matching information for all the matching options.

Given this matching information, the **Driver** and **Passenger** will decide if they want to accept the new matching options. The **Service Provider** will need to modify the **Schedules** based on the user's responses, assign the finalized **Schedules** to the **Driver** agents, and update the **Passenger** agents. Note that, only the scheduled **Itinerary** is created for each **Trip** in the **Schedule**, while the actual **Itinerary** will be modified by the **Driver** agent once receives the **Schedule**.

### 1.2. Simulation flow

The proposed ridesharing simulation platform is an event-based time-dependent model. An event register is designed to sort and send the events in a chronological order. After receiving an event, the simulation clock is progressed to the next event time and execute the event with the relevant agents. We show the designed simulation flow is shown in SI Figure 3.

During initialization of the simulation run, the time-dependent passenger demand and driver supply origin-destination matrices, and agent characteristic information are read to register the *Agent Generation Events*. Meanwhile, the Service Provider agent is created with Operational Module, which imports different operational strategies. The initial *Ridesharing Matching Event* is registered in the event queue for the Service Provider agent.

<u>*Agent Generation Event*</u>

When an *Agent Generation Event* is activated, a batch of Passenger/Driver agents are generated. First, each agent initializes its agent and trip characteristics, and preferred Service Providers. Then, it creates two Stops that correspond to the agent's origin and destination as defined in the trip characteristic attributes. At this stage, all agents are by default *Offline* (not yet send ridesharing requests). Next, an *Online-Offline Events* with their desired online times are registered to trigger ridesharing request. After that, an *Agent Staying/Leaving Decision Making Event* is added to the event queue, indicating the time that the agent might consider quitting the service. Lastly, Driver agents need to initialize their Schedules by calling the Trip Routing module. The Trip Routing module creates a Trip with the Driver's origin and destination and generates two Itineraries (as explained in the previous section). After initializing Driver's Schedule, an *Agent Termination Event* that corresponds to the Driver arriving at the end of the Itinerary (i.e., destination) without any matching is registered as well.

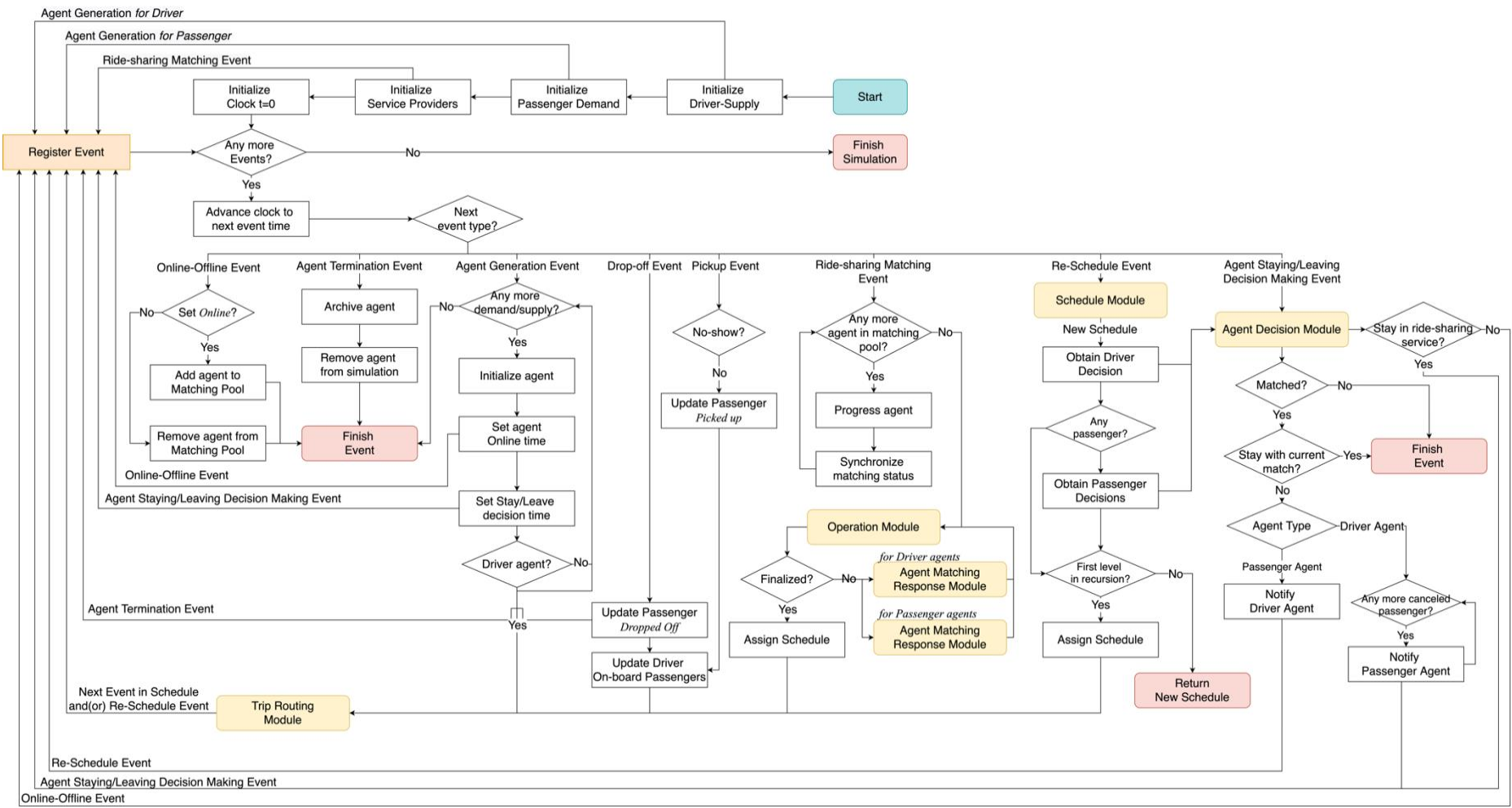

SI Figure 3 Flowchart for ridesharing simulation process

*Online-Offline Event*

When the *Online-Offline Event* is activated, an agent who is requesting a shared ride (*Online*) will be added to the matching pool. Or an agent might not request a shared ride, in this case, he/she will be removed from the matching pool (*Offline*). The matching pool records all the Passengers and Drivers who are waiting for a matching, and the Service Provider will only perform ridesharing matchings between Passengers and Drivers in the matching pool.

After the *Online-Offline Event*s, the simulation platform has initialized some Passenger and Driver agents, which were added to the matching pool. Meanwhile, the simulation has several events in the event queue: *Agent Staying/Leaving Decision Making Events*, *Agent Termination Events*, *Online-Offline Events*, *Agent Generation Events*, and a *Ridesharing Matching Event*.

*Ridesharing Matching Event*

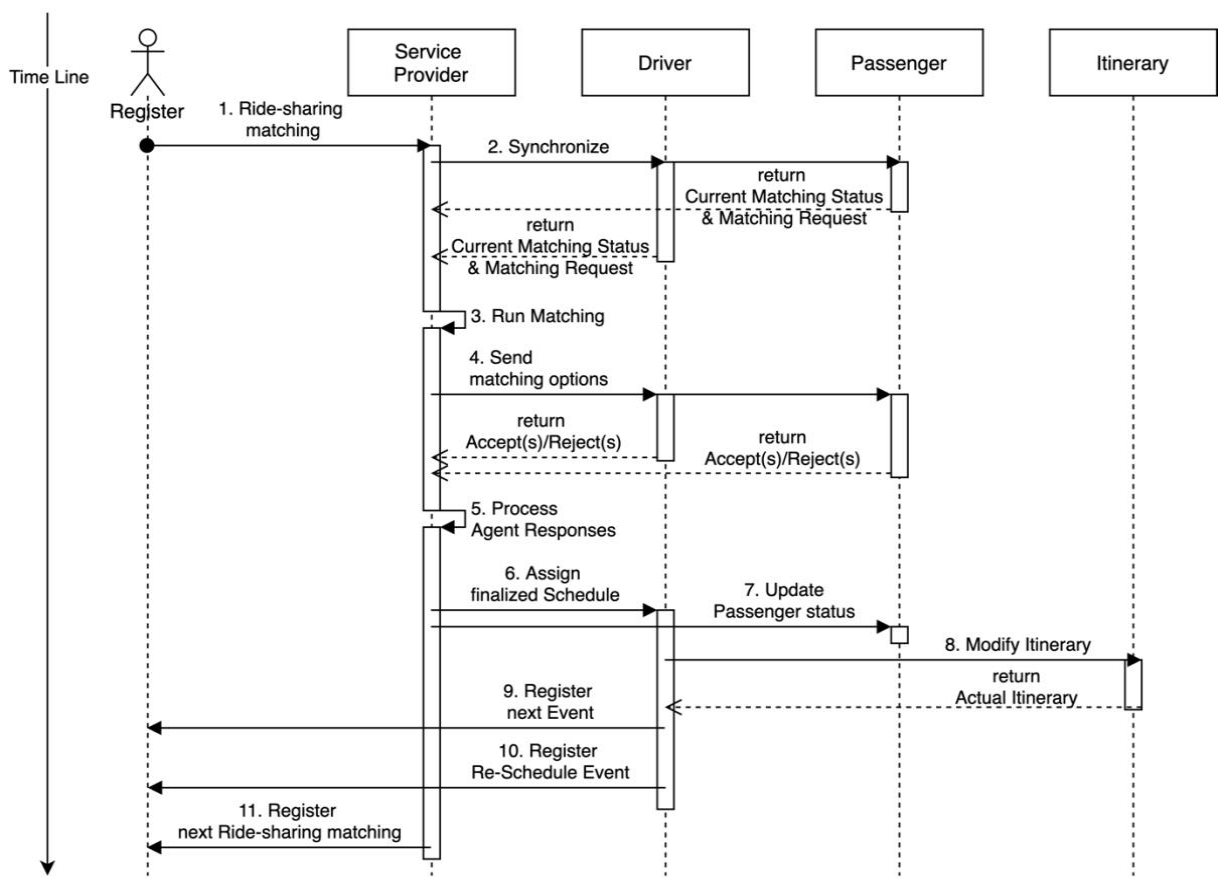

SI Figure 4 Sequence diagram for *Ridesharing Matching Event* routine

The execution of *Ridesharing Matching Event* is illustrated in a sequence diagram (SI Figure 4). During *Ridesharing Matching Event* execution, the Passenger and Driver agents first progress to the current clock and synchronize their current matching status (peer Driver and Passengers, updated Schedules), and ridesharing requests (current location, updated time constraints) with the Service Providers.

The simulation selects Driver and Passenger agents for each service provider from the matching pool, based on the service provider flags. Then, the Service Providers perform ridesharing matching between these Drivers and Passengers using Operation Module and provide information on the matching options based on their operational strategies. This matching option information is sent to all relevant agents, and the Agent Matching Response Module is called to provide agent accept/reject decisions. These agent responses are sent back to the Service Providers, and modifications on the matching options (Schedules) are performed to fulfill the agent requirements.

After that, the finalized Schedule is sent to the Driver, and the Passenger agent status are updated to be *matched waiting for pickup* by the matched Driver. The Driver is set to perform the first Trip in the Schedule with the actual Itinerary and register one of the possible following events: *Pickup Event*, *Drop-Off Event*, or *Agent Termination Events*, with the actual event time. We assume that a driver-specific (stochastic) factor is applied on the average network travel times, to obtain the actual Itinerary and simulate driver-specific travel times.

In case that the next event time in the actual Itinerary is different from the one in the scheduled Itinerary (for a certain threshold that accounts for small variations in stochastic driving speeds, and floating-point errors), a *Re-schedule Event* will be registered in the event queue. This event occurs at the time that the Service Provider could notice that the Driver will not be able to execute the next event according to the scheduled Itinerary. Finally, if there is any remaining event in the event queue, another *Ridesharing Matching Event* will be scheduled at next matching window.

*Agent Staying/Leaving Decision Making Event*

When the *Agent Staying/Leaving Decision Making Event* is activated, an Agent Decision Module is called and returns three possible outcomes: a) stay with current matching, b) leave current matching but still use the ridesharing service, or c) leave the ridesharing service. If the Driver or Passenger still uses ridesharing service (outcome a or b), another *Agent Staying/Leaving Decision Making Event* will be registered. This registered event ensures that Passenger agents who are not matched will eventually be removed from the simulation.

If a Driver or Passenger chooses to leave the current matching (outcome b, SI Figure 5) or even leave the ridesharing service (outcome c, SI Figure 6), there will be some complex interactions between the agents.

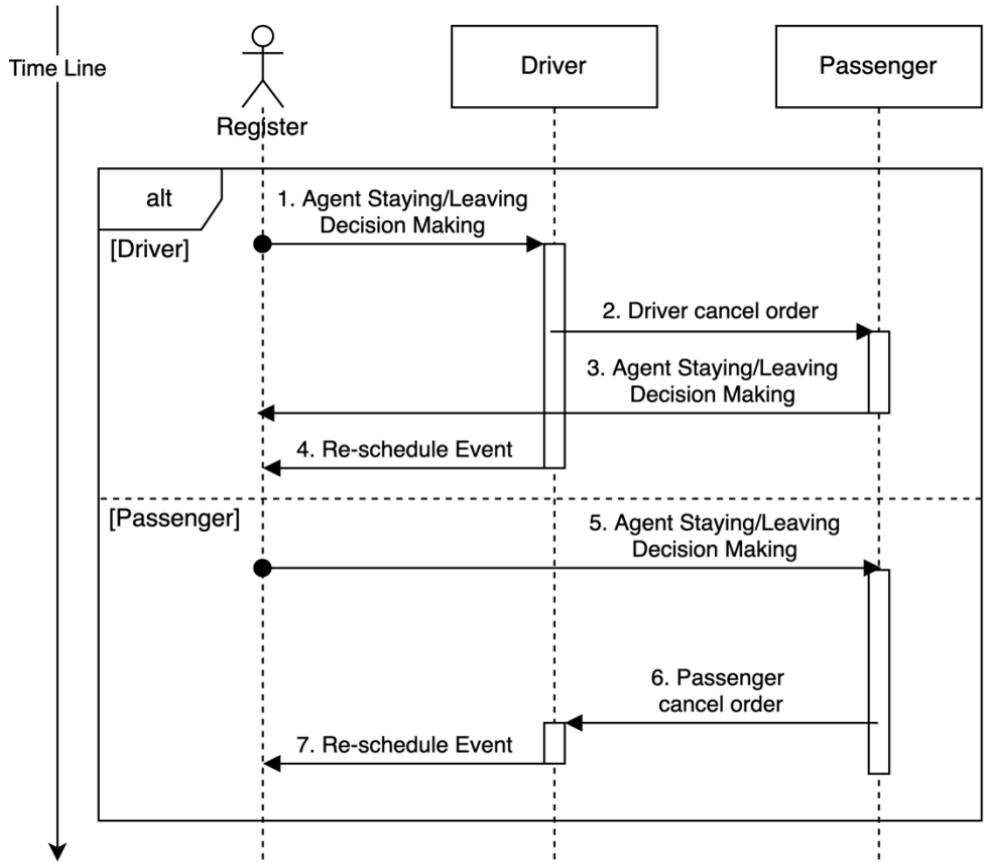

SI Figure 5 Sequence diagram for *Agent Staying/Leaving Decision Making Event* routine of decision outcome (b)

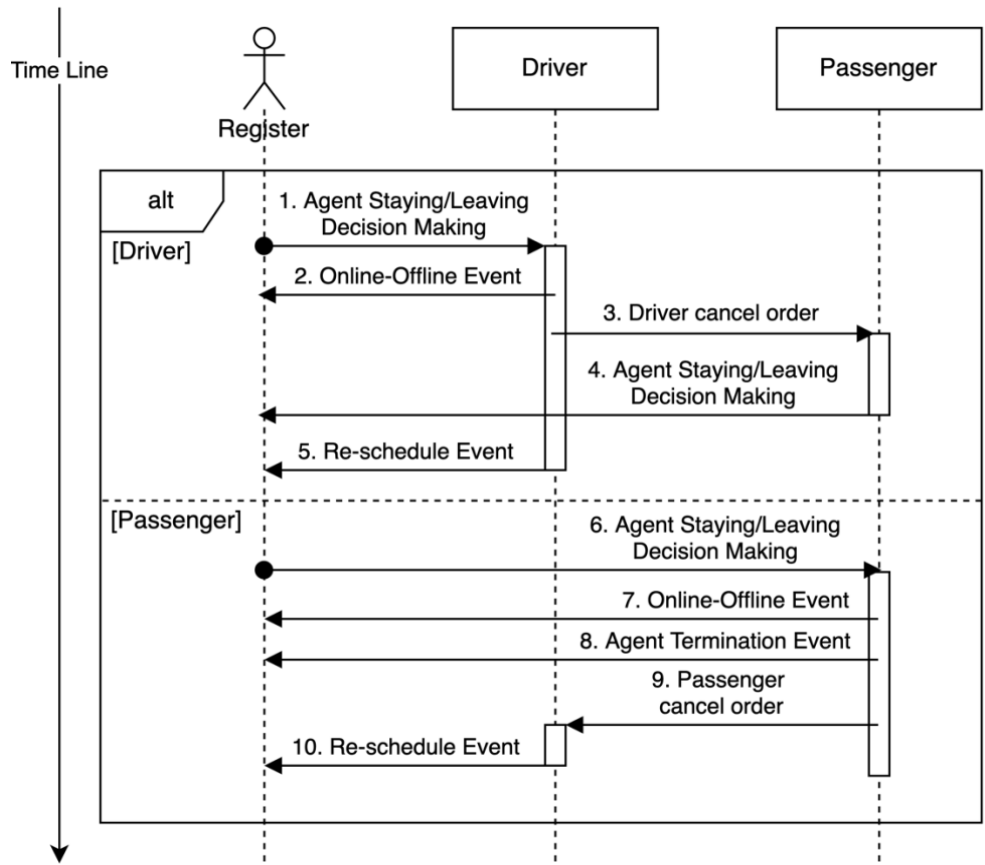

SI Figure 6 Sequence diagram for *Agent Staying/Leaving Decision Making Event* routine of decision outcome (c)

As shown in SI Figure 5, in case Driver agents decide to cancel existing orders, their canceled Passenger agents will be notified. For these Passenger agents, *Agent Staying/Leaving Decision Making Events* will be registered to determine if they are willing to stay in the ridesharing service and wait for another matching. If the Driver agents decide to leave the ridesharing service (outcome c, SI Figure 6), an *Online-Offline Event* will be registered for removing the Driver from matching pool (if assuming he/she will not participate in ridesharing anymore). Lastly, a *Re-schedule Event* is registered for obtaining the Driver's new Schedule with some Passengers excluded.

In case Passenger agents want to cancel the confirmed orders, the corresponding matched Driver will be notified about these cancellations and trigger *Re-schedule Events* for updating the Driver agents' Schedule (SI Figure 5). In case the Passenger agents want to quit the ridesharing services, *Online-Offline Events* will be registered to remove the Passenger agents from the matching pool, and the Passenger agents will be removed from the simulation with an *Agent Termination Event* (SI Figure 6).

<u>Re-schedule Event</u>

As described above, Schedule updates are often required after running *Agent Staying/Leaving Decision Making Event*. We illustrate the simulation routine of *Re-schedule Event* and show the interactions between *Agent Staying/Leaving Decision Making Events* and *Re-schedule Events* in SI Figure 7.

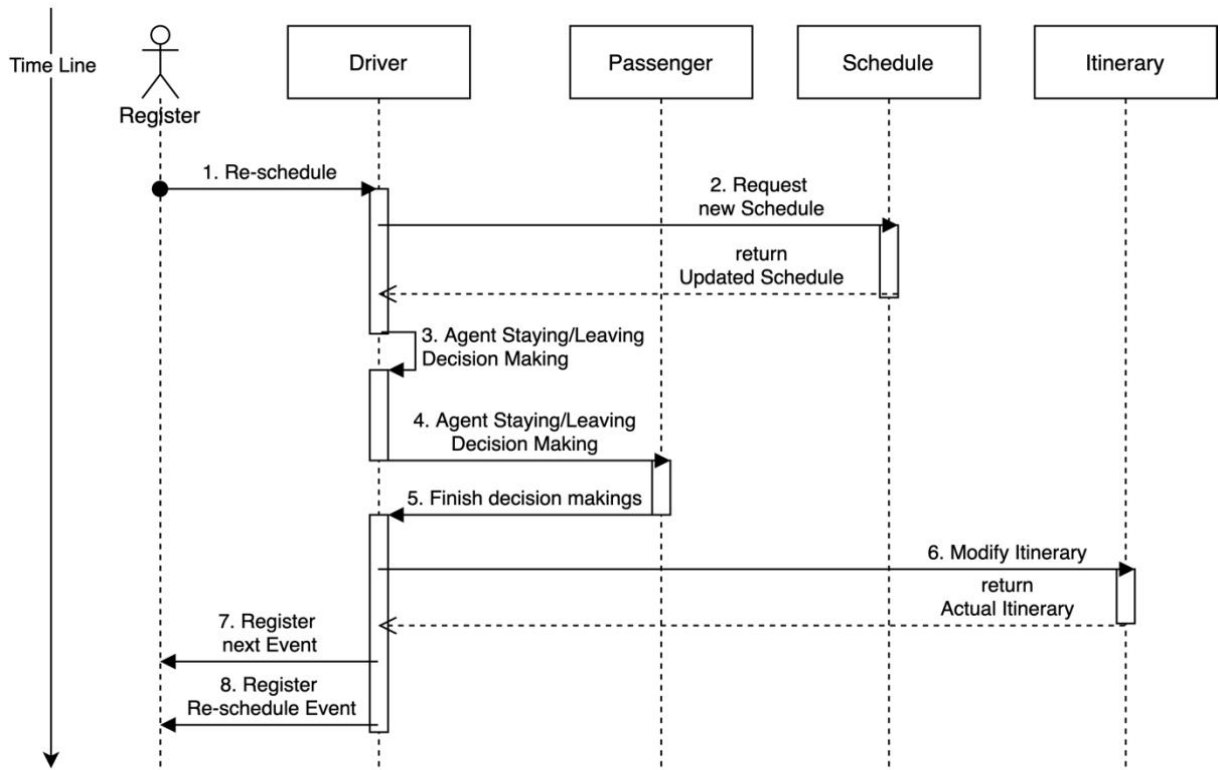

SI Figure 7 Sequence diagram for *Re-schedule Event* routine

When a *Re-schedule Event* is activated, either the schedule or the scheduled event time is updated. For example, some Passengers may cancel their orders, or there is a delay in executing any event. A Schedule Module is called to compute the new Schedule with these updates. *Agent Staying/Leaving Decision Making Events* are executed to determine the Driver's and Passengers' decisions with respect to the updated Schedule. Recall that in SI Figure 6, if Driver or Passenger agents are not satisfied with the new Schedule, the order will be cancelled and in turn the *Re-schedule Event* will be registered again. These recursive decision makings will be finalized until all agents remaining in the Schedule are satisfied. Then, this finalized Schedule will be assigned to the Driver, and the updated matching information will the sent to the corresponding Passengers. Logical checks are introduced in the process to avoid agent inconsistent decisions on the same update in the recursive operations, and to prevent recursive updates on the events.

*Additional Events*

When a *Pickup Event* is activated, the Driver's current location is updated to the pickup location, and his/her capacity is updated accordingly. In case the Passenger does not show up at the pickup location, the Driver's capacity will not be updated, a *Re-schedule Event* will be registered for the Driver. Otherwise, one of the following events is registered: *Pickup Event*, *Drop-Off Event*, or *Agent Termination Event*. Similarly, for a *Drop-Off Event*, the Driver's current location and capacity will be updated, apart from the next event, *Agent Termination Event* for the dropped off Passenger will also be registered.

When *Agent Termination Event* is called for Passengers, this means the Passenger is either being dropped off at his/her destination or has left the ridesharing system, and it is removed from the simulation system. For Driver agents, the *Agent Termination Event* is only called when they reach the destination, no matter whether they have been matched or not.

The simulation then continues getting events from the event queue and terminates when the event queue is empty.

**1.3.    Simulation modules**

We mentioned several modules that play important roles in the simulation. In this subsection, these modules are described. These additional simulation modules represent detailed operation of ridesharing systems, which include: providing routing information for Drivers and tracking their locations for matching, finding Schedules for a set of Passengers, matching Passengers and Drivers, simulating Driver's and Passenger's responses on matching options, assigning finalized matching Schedule, and simulating Driver and Passenger making decisions on order cancellations. These modules enrich the explanatory power of the simulation model and are expected to provide additional capabilities for analyzing and evaluating ridesharing services.

*1.3.1.   Trip Routing Module*

Before generating Itineraries, the Trip Routing Module first updates the Driver's current location by iterating through Driver's actual Itinerary timestamps and finds the next nearest road node as new starting Stop. We illustrate the Driver location update process in SI Figure 8.

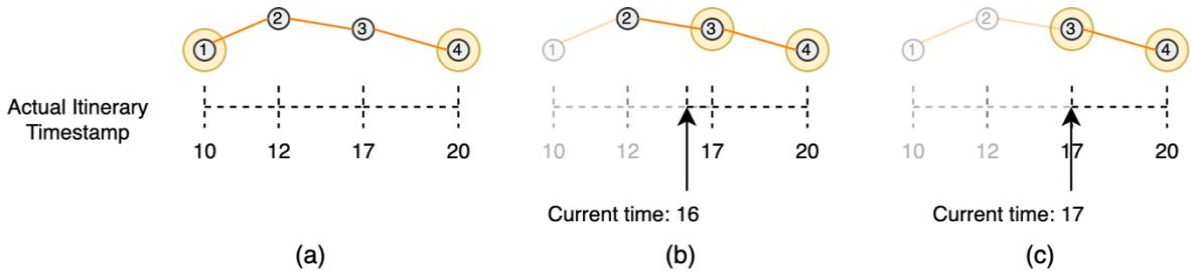

SI Figure 8 Driver location update example

Suppose the current time now is 16, as shown in SI Figure 8(b). The Driver is still traveling on link (2, 3). We assume pickup/drop-off will not be executed in the link. The Driver's earliest possible pickup location is at node 3, which corresponds to time 17. Therefore, the ride-sharing matching will be performed for the Driver as if he/she is at node 3 at time 17 (SI Figure 8c).

We compute the Itineraries using A* shortest path algorithm (Goldberg and Harrelson, 2005), with a modification to handle time-dependent travel times (Dean, 2004) in which Drivers are assumed to travel along links in a First-In-First-Out manner.

*1.3.2.   Scheduling Module*

Given a set of Stops, the Schedule Module calls the Trip Routing Module to obtain the Stop-to-Stop travel times and updates the Driver's current Stop. Together with constraints associated with each Stop (e.g., spatiotemporal constraints, requested seats, etc.), the Schedule Module finds a Stop sequence that satisfies these constraints, which is the well-known dial-a-ride problem (DARP). We implemented an efficient DARP algorithm with Dynamic Tree Structure (Yao and Bekhor, 2021) for finding these Schedules.

*1.3.3. Operational Module*

The Operational Module is the core of Service Providers. It deals with three main operations: matching Passengers and Drivers, setting the price for the trips, and processing Passenger and Driver's acceptance/rejection responses to the matching options. This subsection focuses on processing agent responses to matching options.

After allocating the costs for all the matched Driver-Passengers combinations, all matched Drivers and Passengers receive cost-earning information, and the Schedules are sent to the Drivers. The acceptance/rejection decisions of Drivers and Passengers are then provided to the Service Providers. We employ a majority-voting scheme for finalizing the matching options as shown in SI Figure 9. The matching options that include new Passengers are first sent to the Driver and all previously matched Passengers. If the majority (of the Driver and previously matched Passengers) agrees to accept the new matching option with the new Passengers, these new Passengers can then decide whether to accept the matching option. If all the new Passengers agree, then the new matching option is finalized. In case there are even votes, the Driver has the final decision for the already matched passengers (and asks for new passengers' decisions), and new passengers can still reject the matching.

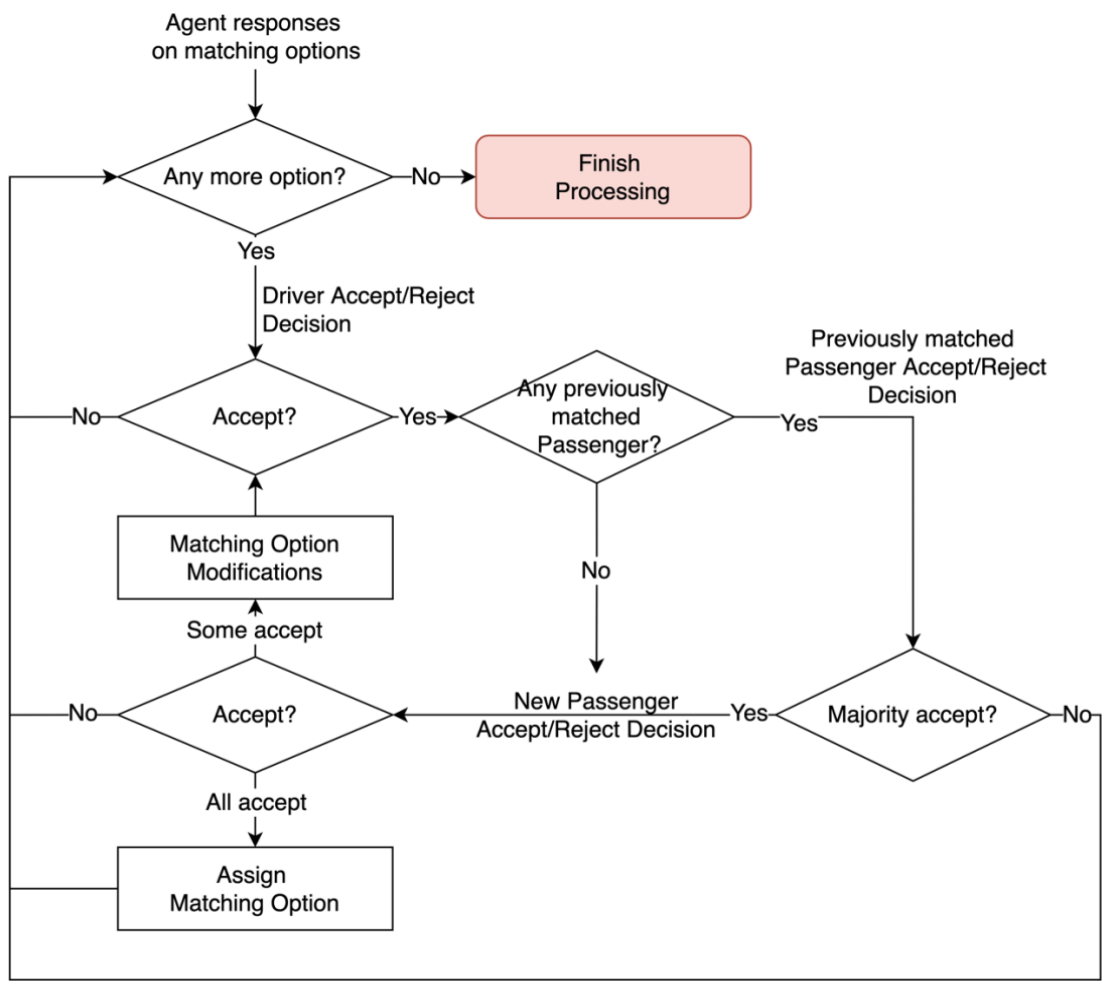

SI Figure 9 Service Provider processing agent accept/reject responses

If some (but not all) of the new passengers accept the new matching, a heuristic approach is adapted to finalize the matching options. The new passengers are sorted by acceptance probability, and new matching options are updated by removing the new passenger with lowest acceptance probability one at a time. Accordingly, the Schedules of the updated matching option are re-computed using the *Scheduling Module*, and the cost-earning information is updated as well. For each updated matching option, all remaining agents are asked about acceptance/rejection decisions. This process stops until the driver and all passengers are satisfied with (or reject) the updated matching option.

## 2. RIDESHARING PRICING BOUNDS

### 2.1. Ridesharing passenger pricing bounds

Similar to the derivation of the ridesharing driver pricing bounds, we derive in the following pricing bounds for ridesharing passengers. For simplicity, we also assume the value of time of travel-alone passengers, $VOT_{rv_j}$, represents the combined monetary cost of travel time and operational costs, while only monetary cost of travel time is considered in the value of time of ridesharing passenger, $VOT_{r_j}$. Following these assumptions, the utility of rejecting new matching options and travel alone $U_{r_j,0}$ for passenger $r_j$ can be written as:

$$U_{r_j,0} = \beta_{time,r_j} \cdot tt_r(r_j) + \beta_{cost,r_j} \cdot cost(r_j)$$
$$= -VOT_{v_i} \cdot tt_r(r_j) \tag{SI 1}$$

and the utility of staying in ridesharing for passenger $r_j$, $U_{r_j,1}$ can be written as:

$$U_{r_j,1} = \beta_{time,r_j} \cdot [tt_r(r_j) + wt(r_j)] + \beta_{cost,r_j} \cdot [\beta_{\exp pay,r_j} \cdot cost(r_j)]$$
$$= -VOT_{r_j} \cdot [tt_r(r_j) + wt(r_j)] - \beta_{\exp pay,r_j} \cdot cost(r_j)$$
$$= -VOT_{r_j} \cdot [tt_r(r_j) + wt(r_j)] - p \cdot d(m) \tag{SI 2}$$

where, $p$ is the cost/km, and $d(m)$ is the payable (shared) distance of matching option $m$ for passenger $r_j$, which can be expressed in terms of average speed $s(m)$, total trip time $[tt_r(r_j) + wt(r_j)]$, and average (weighted) payable shared percentage $\alpha(r_j)$:

$$d(m) = \alpha(r_j) \cdot s(m) \cdot [tt_r(r_j) + wt(r_j)] \tag{SI 3}$$

To have more passengers staying for new matchings, a naïve approach will be finding $p$, such that the utility of (quitting) travel-alone to be lower than staying in ridesharing:

$$U_{r_j,1} = -VOT_{r_j} \cdot [tt_r(r_j) + wt(r_j)] - p \cdot \alpha(r_j) \cdot s(m) \cdot [tt_r(r_j) + wt(r_j)]$$
$$\geq U_{r_j,0} = -VOT_{v_i} \cdot tt_r(r_j) \tag{SI 4}$$

and the upper bound on price for passenger $r_j$ is:

$$p \leq \frac{\left(VOT_{rv_j} - VOT_{r_j}\right) \cdot tt_r(r_j) - VOT_{r_j} \cdot wt(r_j)}{\alpha(r_j) \cdot s(m) \cdot \left(tt_r(r_j) + wt(r_j)\right)} \tag{SI 5}$$

## 2.2. Toy network example for pricing bounds

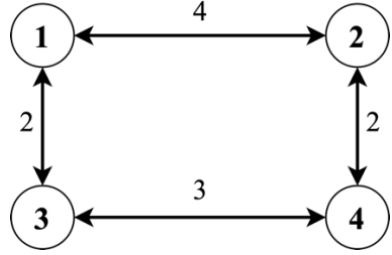

SI Figure 10 Toy network with link travel times [min]

We illustrate pricing bounds using a toy network example (SI Figure 10), in which drivers (SI Table 2) and passengers (SI Table 3) request ridesharing. For simplicity, we assume each driver has capacity of 1, and the matchings are shown in SI Table 1, where the average billable percentages $\alpha(m)$ and payable percentages $\alpha(r_j)$ are calculated, and $s(m)$ is obtained from the network.

SI Table 1 Matching options for the toy network

| Matching option $m$ | Driver Route | $\alpha(m), \alpha(r_j)$ | $s(m)$ |
|---|---|---|---|
| $(v_1, r_1)$ | $1 \to 3 \to 4$ | $\frac{5}{5}, \frac{5}{5}$ | $60[km/hr]$ |
| $(v_2, r_2)$ | $1 \to 3 \to 4$ | $\frac{5}{5}, \frac{5}{5}$ | $60[km/hr]$ |
| $(v_3, r_3)$ | $1 \to 3 \to 4$ | $\frac{5}{5}, \frac{5}{5}$ | $60[km/hr]$ |
| $(v_4, r_4)$ | $1 \to 2 \to 4$ | $\frac{2}{6}, \frac{2}{2}$ | $60[km/hr]$ |
| $(v_5, r_5)$ | $1 \to 3 \to 4$ | $\frac{5}{5}, \frac{5}{5}$ | $60[km/hr]$ |

SI Table 2 Example of pricing bounds for driver agents

| Driver ID $v_i$ | (origin, destination), $tt_v(v_i)$ | $VOT_{v_i}$ | $\Delta t(v_i)$ | $\underline{p_{v_i}}$ | $\overline{p_{v_i}}$ |
|---|---|---|---|---|---|
| $v_1$ | (1, 4), 5 | 4 | 3 | 1.5 | **4** |
| $v_2$ | (1, 4), 5 | 4 | 5 | 2 | 4 |
| $v_3$ | (1, 4), 5 | 5 | 5 | 2.5 | 5 |
| $v_4$ | (1, 4), 2 | 3 | 3 | **3** | 9 |
| $v_5$ | (1, 4), 5 | 4 | 3 | 1.5 | 4 |

By using eq. (23), we compute the lower and upper bounds for each driver agent $v_i$. If we would like to attract all drivers, the price $p$ should be set according to eq. (25) as:

$$\max_{v_i} \underline{p_{v_i}} = 3 \leq p \leq 4 = \min_{v_i} \overline{p_{v_i}}$$

SI Table 3 Example of pricing bounds for passenger agents

| Passenger ID $r_j$ | (origin, destination), $tt_r(r_j)$ | $VOT_{r_j}$ | $VOT_{rv_j}$ | $wt(r_j)$ | $\overline{p_{r_j}}$ |
|---|---|---|---|---|---|
| $r_1$ | (1, 4), 5 | 1 | 6 | 1 | $\overline{4}$ |
| $r_2$ | (1, 4), 5 | 1 | 6 | 0 | 5 |
| $r_3$ | (1, 4), 5 | 0 | 6 | 1 | 5 |
| $r_4$ | (2, 4), 2 | 1 | 6 | 0 | 5 |
| $r_5$* | (1, 4), 5 | 1 | 3 | 1 | $\overline{1.5}$* |

Similarly, by using eq. (26), the upper bounds for each passenger agent $r_j$ are reported in SI Table 3. If we would like to attract all passengers, the price $p$ should be set according to eq. (30) as:

$$0 \leq p \leq 1.5 = \min_{r_j} \overline{p_{r_j}}$$

However, since $\min_{r_j} \overline{p_{r_j}} = 1.5 < \max_{v_i} \underline{p_{v_i}} = 3$ violates eq. (31), we could not design a pricing scheme that attracts all passengers and drivers. Alternatively, if we do not match passenger $r_5$ with driver $v_5$ (marked with star), there exists a feasible pricing scheme to attract the rest of the matched passengers and drivers.

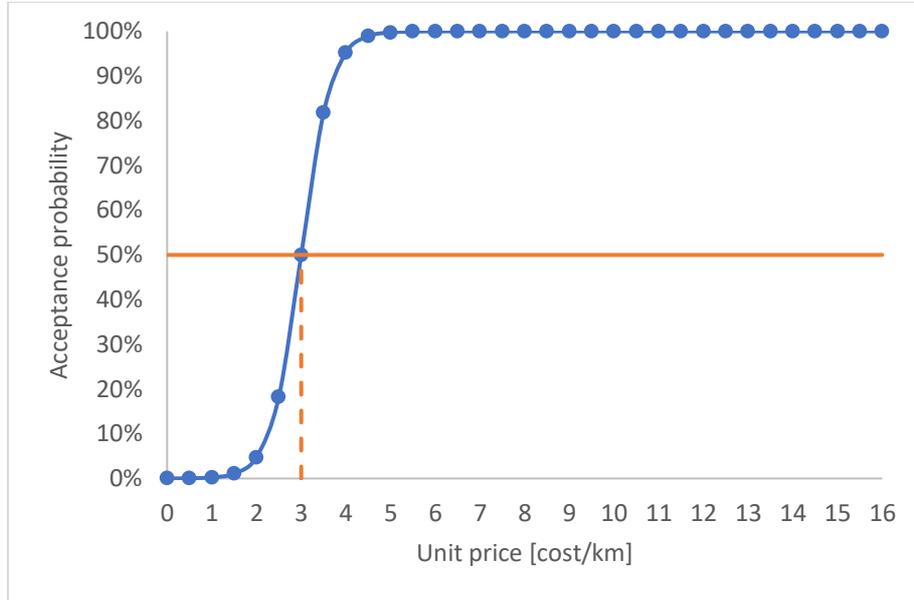

SI Figure 11 Example of matching option acceptance probability

We demonstrate in SI Figure 11 the matching option acceptance probability, in the case that driver $v_4$ only chooses between accept the matching option, or reject and travel alone (i.e., binary logit). The driver's price lower bound can be obtained when the acceptance probability is 50% ($U_{v_i,m} = U_{v_i,0}$), which is $\underline{p_{v_i}} = 3$. Note that, these analytical bounds can only be obtained in very simple cases, due to the complex interactions between the parameters and matching variables ($\alpha(m), s(m)$) in large real-size problems.

## 3. EXPERIMENT RESULTS

In this section, we show time-series results in status profiles for Passengers (SI Figure 12), Drivers (SI Figure 13) and the Service Provider (SI Figure 14). The performance metrics of the experiment are calculated over 10 replications. Note that, two periods, each of 10% extra supply and demand, are added as warmups to the simulation.

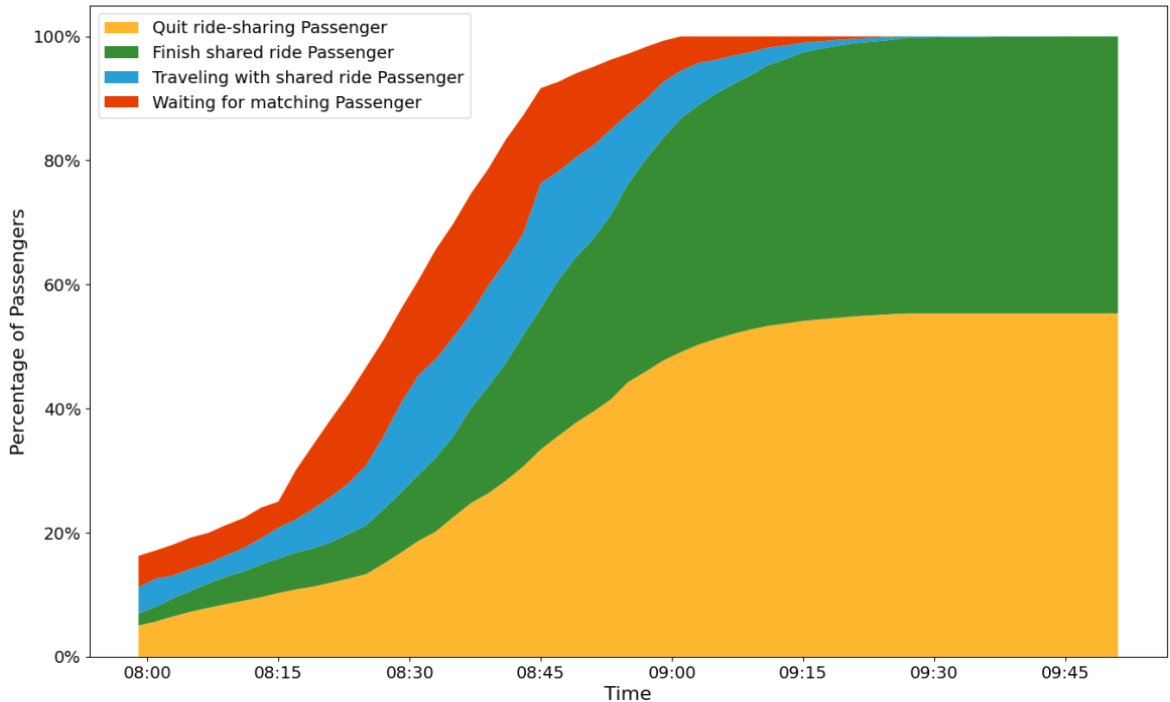

SI Figure 12 Passenger status profile

The Passenger status profile shows the simulation behavior is logical. Most of the Passenger demands appear in the simulation between 8:15-8:45, following our specified demand pattern. As the simulation progresses, ridesharing traveling Passengers and waiting for matching Passenger gradually decrease, and their status are switched as Finished shared ride or Quitted ridesharing service and removed from the simulation system. The number of waiting for matching Passengers remains almost constant between 8:30-8:45, which is related to high demand during this period. With this demand, the assumed ridesharing system could be oversaturated, or no matching could be found for these Passengers which satisfy their spatiotemporal constraints.

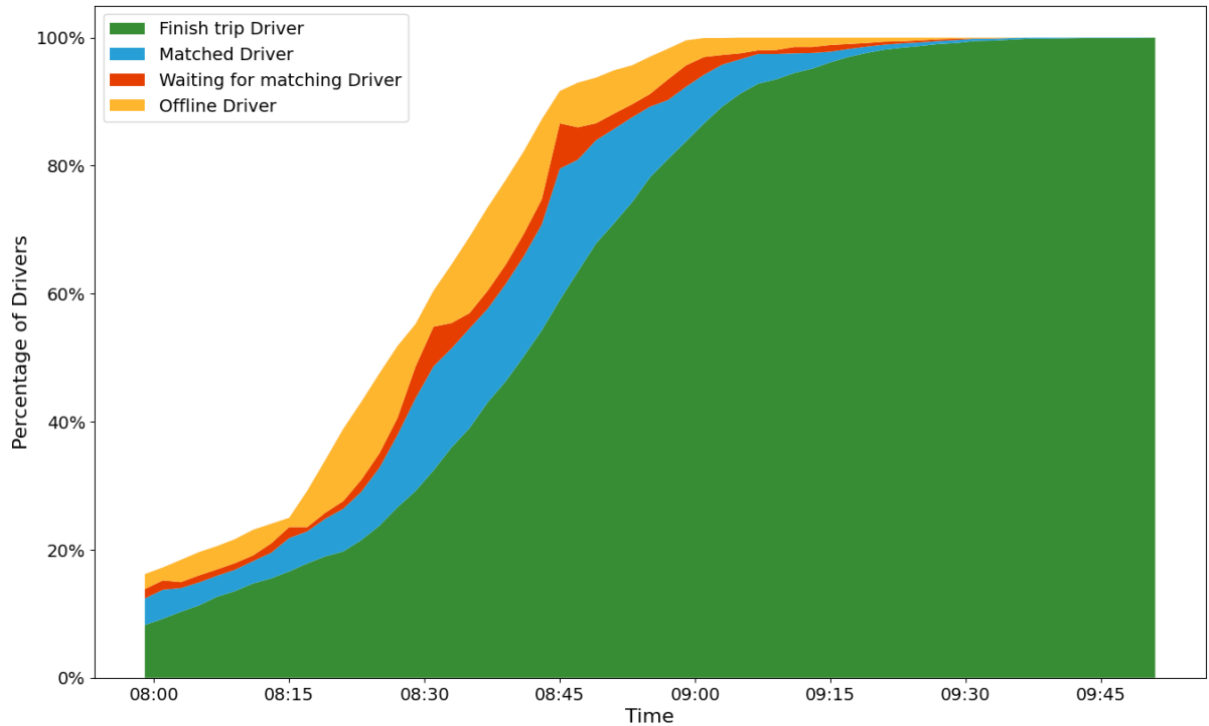

SI Figure 13 Driver status profile

As shown in SI Figure 13, most of the Driver supplies are generated during 8:15-8:45 as specified. At the end of the simulation, all the Drivers finish their trips, and are removed from the simulation. For Drivers who are still traveling in the network (and remained in simulation), his/her status fails in one of the three status: Matched, Waiting for matching, or Offline. These traveling Drivers gradually arrive at their destinations and switch to Finish trip. The increase of number of Drivers finishing their trips, is almost the same as the increase of Drivers. One explanation is that the Drivers' trip length is too short. As shown in SI Figure 10 in the main text, the average trip length associated with the OD demand pattern is 10.46 km, and the average network link speed is 49.45 km/hr, which results average travel alone trip duration to be 12.7 min. The average travel alone trip duration is shorter than the OD demand period length 15 min, which explains why the Drivers are quickly switched to Finish trip. This result is consistent with literature (Agatz et al., 2011; Stiglic et al., 2016), in which the success rate is the lowest for short trips.

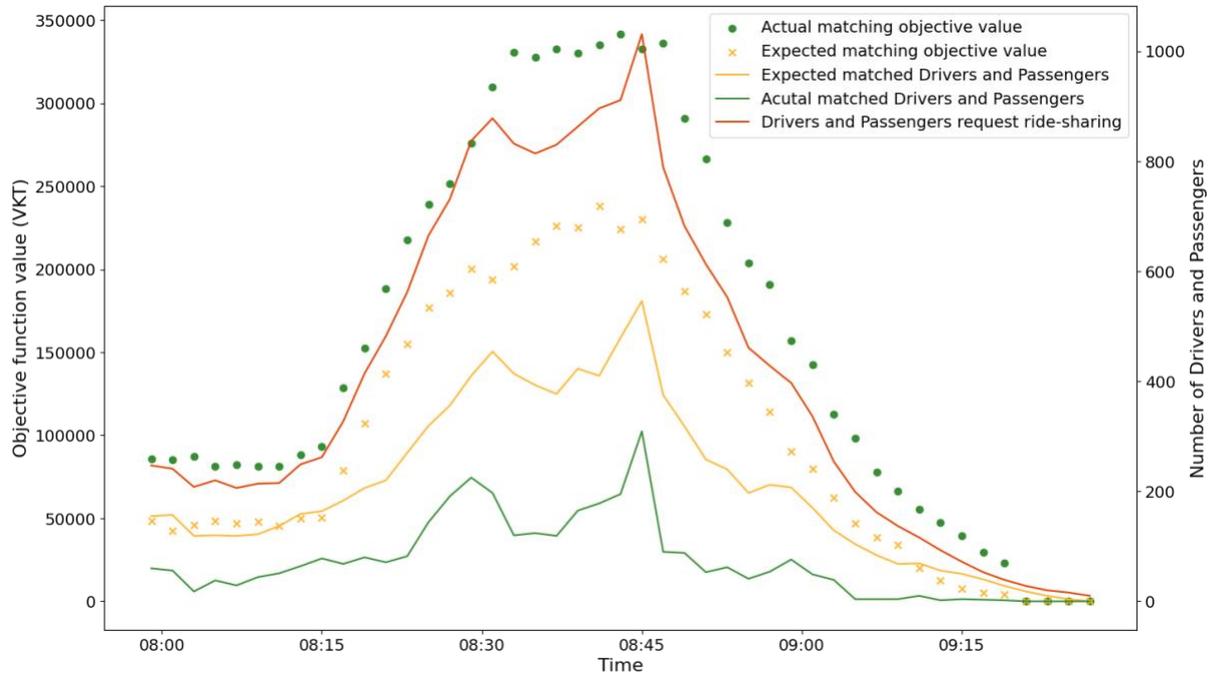

SI Figure 14 Service Provider performance profile

We show the Service Provider performance profile in SI Figure 14, in which the number of requested, expected matched, and actual matched Drivers and Passengers are compared. During 8:30-8:45, there is a spike in number of Driver and Passenger ridesharing requests due to the specified OD demand pattern. However, the Service Provider could not find matches (gap between red and orange line) that satisfy the Driver and Passenger's spatiotemporal constraints and results in Drivers and Passengers waiting for matchings.

Furthermore, only part of the found matchings (expected matching) could be finalized (actual matching) for the Drivers and Passengers. This difference in expected matchings and actual matchings will create a gap between expected and actual matching objective value. The actual objective function value is found by calculating the VKT of the actual matched and not matched Drivers and Passengers using equation (1). Remind that our matching objective is to minimize the overall VKT, where larger value means poorer matching performance. As shown in SI Figure 14, the actual objective value is larger than the expected one almost during all the simulation. This suggests the Service Provider is overestimating the performance of the ridesharing matching. We show in this experiment that our simulation platform has the potential for ridesharing system analysis and improvement for service providers and could share additional insights for regulators.

## 4. SENSITIVITY ANALYSIS ON DRIVER SUPPLY LEVEL

Table 4 Selected factors and their levels for the sensitivity analysis

| Factor | Default Level | Analysis Levels |
|---|---|---|
| Driver supply | 50% | 1%, 5%, 10%, 20%, 40%, 50%, 60%, 80%, 100%, 120%, 150%, 200%, 250%, 300%, 350%, 400% of the Passenger demand (2,000 Passengers) |

We show the performance metrics with respect to different levels of Driver supply in SI Figure 15. The impact on Passenger and Driver excess travel time is shown in SI Figure 15(a), in which both Driver and Passenger excess travel time decrease as number of Drivers increase. This result is expected since, with more Drivers, it is more likely the ridesharing Service Provider could match Drivers and Passengers with similar trip characteristics (origin, destination, time constraints), therefore reduce the excess travel times.

The impacts on Drivers, Passenger, and overall success rate are shown in SI Figure 15(b). With increase in number of Drivers, the Passenger success rate (i.e., percentage of Passengers who finish their trips with ridesharing) increases. This is not only related to the greater supply of Drivers providing shared rides; the Drivers are spread around in the network and can reach the Passengers much faster (like taxi). This also explains the Passenger expected pickup time monotonically decrease as Driver supply level increase (SI Figure 15(c)).

We observe that the Driver success rate starts to increase at low number of Drivers and reaches maximum Driver success rate at around 800 Drivers (supply level at 40%), after which the Driver success rate starts to decrease. This is because Drivers are scattered around in the network at low Driver supply level, and the matching with a Passenger is highly dependent on the spatiotemporal constraints. When the Driver supply level is too high and there is not enough Passenger demand, many Drivers cannot find a matching and results in decrease of Driver success rate.

The overall success rate captures the adverse effect of Driver supply level in Driver and Passenger success rates. With low level Driver supply, the Passenger success rate dominates the overall success rate, which results in low overall success rate. As the number of Driver increases, the effect of low Driver success rate kicks in, and lowers the overall success rate. We observe the maximum overall success rate is reached at around 2,000 Drivers, which corresponds to 100% supply level (of Passenger demand). This may suggest that to make ridesharing work, we need to have a balanced supply and demand. This result is consistent with one of the common objectives in the literature: driver supply and passenger demand balance (Wang and Yang, 2019).

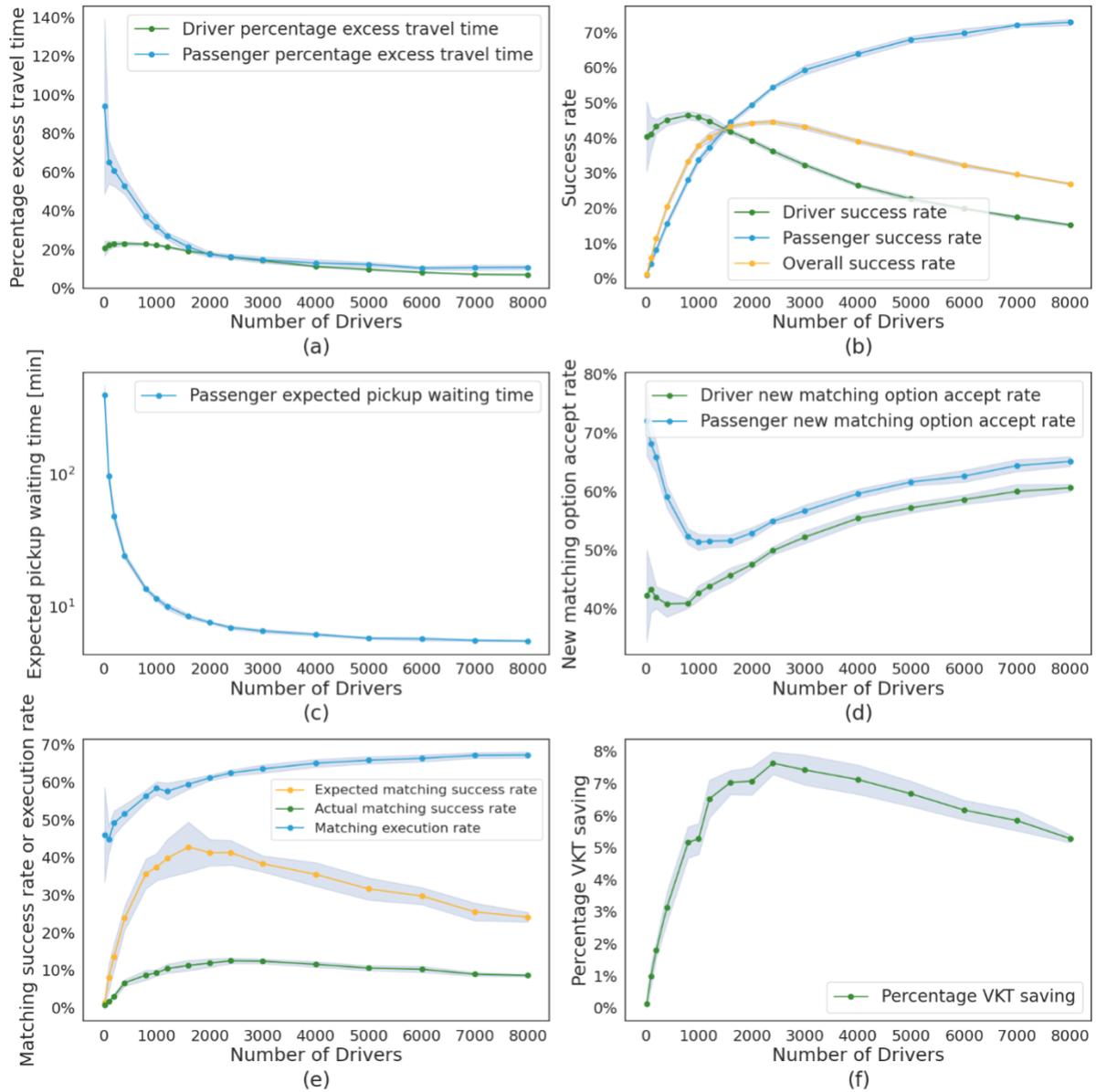

SI Figure 15 Performance metrics for demand sensitivity analysis

In SI Figure 15(d), we show new matching option accept rate for Drivers and Passengers. Remind that, the new matching option accept rate is only calculated for Drivers and Passengers who receive new matching options. Therefore, only few Drivers and Passengers will receive matching options at low Driver supply level, and results in high variance in accept rates. The Passenger accept rate starts to decrease at low number of Drivers, and reaches minimum Passenger accept rate at around 1,000 Drivers (supply level at 50%), after which the Passenger accept rate starts to increase. The decreases in Passenger accept rate can be explained that, as the number of Driver increases, more Passengers are receiving matching options, but the level of Driver supply cannot provide attractive matching options (i.e., matching options at this supply level have large detours and long pickup waiting time), and results in Passenger rejection.

As the Driver supply level increases, the Service Provider could find better matchings with shorter detours and pickup times for the Drivers and Passengers. Therefore, both the Passenger and Driver are satisfied with the new matching options, and their accept rates increase. Moreover, Drivers and Passengers are less likely to cancel matchings at high supply level, and results in increases in matching execution rate (SI Figure 15(e)).

The expected and actual matching success rates are shown in SI Figure 15(e). The expected matching success rate shares the same pattern as the overall success rate, in which it first increases then decreases, and reaches maximum at around 2,000 Passengers. However, even the expected matching success rate start to decrease from 2,000 Passengers, the actual matching success rate is only slightly decreased. This is related to the increases Driver and Passenger new matching accept rate, in which they are more willing to accept the proposed new matching options.

We show the percentage VKT savings in SI Figure 15(f). Since more Passengers could be matched with the Drivers, the percentage VKT savings first increase as Driver level increases and reach maximum at around 2,400 Drivers (120% supply level). After that, the percentage VKT savings gradually decreases. This because the total base-case VKT increases as the number of Drivers increased, and the total number of shared trips (bounded by the fixed number of Passengers) remains the same.